\documentclass[a4paper,11pt,UKenglish]{article}
\usepackage{graphicx}
\usepackage{amssymb}

\usepackage{pifont}

\usepackage{multirow}
\usepackage{centernot}

\newcommand{\pt}{\mbox{$p_{\mathrm{T}}$}}
\newcommand{\pti}{\mbox{$p_{\mathrm{T},i}$}}
\newcommand{\mpt}{\mbox{$\centernot{p\,}\!_{\mathrm{T}}^{\parallel}$}}
\newcommand{\Et}{\mbox{$E_{\mathrm{T}}$}}
\newcommand{\Etl}{\mbox{$E_{\mathrm{T,1}}$}}
\newcommand{\Ets}{\mbox{$E_{\mathrm{T,2}}$}}
\newcommand{\ptjet}{\mbox{$p_{{\mathrm{T}}}^{\mathrm{jet}}$}}
\newcommand{\ptgamma}{\mbox{$p_{{\mathrm{T}}}^{\gamma}$}}
\newcommand{\ptZ}{\mbox{$p_{{\mathrm{T}}}^{Z^0}$}}
\newcommand{\ptparticle}{\mbox{$p_{{\mathrm{T}}}^{\mathrm{particle}}$}}
\newcommand{\AJ}{\mbox{$A_{\mathrm{J}}$}}

\newcommand{\Iaa}{\mbox{$I_{\rm AA}$}}

\newcommand{\Rcpcent}{\mbox{$R_{\rm CP}^{\rm cent}$}}
\newcommand{\Raacent}{\mbox{$R_{\rm AA}^{\rm cent}$}}

\newcommand{\ANcollcent}{\mbox{$\langle N_{\rm coll}^{\rm cent} \rangle$}}
\newcommand{\ANcollperf}{\mbox{$\langle N_{\rm coll}^{\rm 60-80} \rangle$}}
\newcommand{\ANcoll}{\mbox{$\langle N_{\rm coll} \rangle$}}

\newcommand{\sqrtsnn}{\mbox{$\sqrt{s_{\mathrm{NN}}}$}}

\newcommand{\Rcp}{\mbox{$R_{\rm CP}$}}

\newcommand{\Raa}{\mbox{$R_{\rm AA}$}}

\newcommand{\antikt}{\mbox{anti-\kt}}

\newcommand{\kt}{\mbox{$k_\mathrm{T}$}}

\usepackage[bookmarks=false]{hyperref}
\usepackage{url} 

\begin{document}

\markboth{Martin Spousta}
{Jet Quenching at LHC}


\title{JET QUENCHING AT LHC}

\author{Martin Spousta\\
~ \\
{\footnotesize
Institute of Particle and Nuclear Physics,
}
~ \\
{\footnotesize
Charles University in Prague, Czech Republic
}
~ \\
{\footnotesize {\it martin.spousta@cern.ch }
}}


\maketitle


\begin{abstract}

We review up-to-date results on high-\pt\ particles and jets in heavy ion collisions by 
three major LHC experiments, ALICE, ATLAS, and CMS. Results of analyses of 2010 
and 2011 Pb+Pb data at $\sqrtsnn = 2.76$~TeV are discussed. 
  We concentrate mainly on results by fully reconstructed jets and discuss 
similarities and important differences in measurements among experiments. We 
point to the importance of understanding the results in a view of
difference between quark-initiated and gluon-initiated jets.


\end{abstract}


\section{Introduction}

Collisions of heavy ions at ultra-relativistic energies can produce hot and 
dense colored medium where relevant degrees of freedom are not complex hadrons 
but deconfined quarks and gluons~\cite{aaag:Karsch}. This consequence of 
asymptotic freedom of quantum chromodynamics (QCD) allows to investigate 
fundamental properties of strong interaction as originally suggested in 
Refs.~\cite{aaad:Collins,aaae:Cabbibo,aage:Shuryak}.
  High transverse momentum (\pt) partonic interactions in perturbative quantum 
chromodynamics (pQCD) lead to a production of two highly virtual back-to-back 
partons (in the second order of pQCD) which subsequently evolve as parton 
showers, hadronize, and are experimentally observed as back-to-back di-jet 
events in the detector. If the partons traverse on their path a dense colored 
medium they can loose energy. The result of the energy loss can be detected as 
modifications of jet yields and jet properties. This phenomenon is commonly 
referred to as the jet quenching. The jet quenching was first proposed by 
Bjorken~\cite{aaac:Bjorken} as an experimental tool to investigate properties of 
the dense medium.


  This review summarizes experimental results on the jet quenching from three 
major experiments at Large Hadron Collider (LHC), the ALICE~\cite{aaar:ALICE}, 
ATLAS~\cite{baab:ATLAS}, and CMS~\cite{aaaq:CMS} experiment. The LHC heavy ion 
program follows up the successful research in the field of ultra-relativistic 
heavy ion collisions performed at Super Proton Synchrotron (SPS) and later at 
Relativistic Heavy Ion Collider (RHIC). The results of four experiments at RHIC 
-- BRAHMS, PHENIX, PHOBOS, and STAR have been summarized in 
Refs.~\cite{aaap:BRAHMS,aaam:PHENIX,aaao:PHOBOS,aaan:STAR} and further discussed 
along with results from SPS e.g. in 
reviews~\cite{aagh:BraunMunzinger,aagc:Muller,aagf:Jacobs,aagg:Huovinen,aagd:Shuryak,aags:Accardi}. 
The increase of the collisional energy by an order of magnitude at LHC compared 
to RHIC allowed collecting high statistics samples of fully reconstructed 
jets, high-\pt\ photons, and intermediate vector bosons $W^{\pm}$, $Z^{0}$. 
These hard probes can shad a new light on the jet quenching and properties of 
the medium.

   There are handful of experimental results at low and intermediate transverse 
momentum\footnote{Throughout this paper, natural units are applied with 
$c=\hbar=1$.} ($\pt \lesssim 20$~GeV) by all three LHC collaborations such as 
measurements of global event properties, particle correlations, and identified 
particle yields. Also, there are important measurements on quarkonia 
suppression provided by LHC experiments. We will not discuss these results 
since primary focus of this review is the jet quenching. For extensive review 
of these topics at LHC see e.g. Refs.~\cite{aafd:Muller,aafc:Singh}.
 
  The recent development in the theory of the jet quenching is summarized e.g. 
in Refs.~\cite{aaca:Armesto,aafm:Bass} or 
reviews~\cite{aaai:Majumder,aaah:Wiedemann,aadj:MehtarTani}. No unified description
of the jet quenching exists. There are different phenomenological 
approaches~\cite{aahg:Baier,aahb:Gyulassy,aahc:Wang,aahd:Zakharov,aahe:Arnold} 
to the calculation of jet quenching, however the underlying principle remains 
the same: hard partons can loose their energy via gluon radiation (an analogue 
of LMP effect in Quantum Electrodynamics~\cite{aagj:Gyulassy,aagk:Wang}) and/or 
by elastic (collisional) energy loss~\cite{aaac:Bjorken} which is of importance 
namely for the heavy quark energy loss~\cite{aagl:Mustafa}.
  Different models not only 
use different techniques and approximations but also they characterize the medium in 
terms of different primary model parameters. There is therefore a great importance in 
performing experimental measurements of jets and other high-\pt\ probes to help to constrain existing models.
 The first experimental evidence of the jet quenching at LHC has been provided by the asymmetry 
measurement discussed in the next section.

\section{The first evidence of the jet quenching at LHC}

  The first experimental evidence of the jet quenching at LHC has been observed in 
the measurement of the di-jet asymmetry~\cite{baaa:ATLAS,aaab:CMS}. The di-jet 
asymmetry has been defined as
  \begin{equation}
\AJ = \frac{\Etl - \Ets}{\Etl + \Ets}
  \end{equation}
  where $\Etl$ resp. $\Ets$ is the transverse energy of the leading resp. subleading jet 
in the event. 

Energy loss of parent partons in the created matter may reduce or ``suppress'' the 
rate for producing jets at a given \Et. This suppression is expected to increase 
with medium temperature and with increasing path length of the parton in the medium. 
As a result, there should be more suppression in central Pb+Pb collisions 
which have nearly complete overlap between incident nuclei, and little or no 
suppression in peripheral events where the nuclei barely overlap. This was indeed 
observed in the measurement of di-jet asymmetry where the jet suppression exhibits 
itself by an increase of the number of events with larger jet asymmetry compared to 
Monte-Carlo (MC) reference.



This asymmetry has been accompanied by a balance in azimuth, that is jets in 
the di-jet system remain ``back-to-back'' despite to a sizable modification of 
their energy. The original~\cite{baaa:ATLAS,aaab:CMS} and 
updated~\cite{baam:ATLAS,aabk:CMS} experimental observation was followed by 
theoretical 
work~\cite{aacb:Qin,aace:Young,aacf:Renk,aade:He,aadg:CasalderreySolana,aahk:Lokhtin} 
suggesting that the suppression can be
explained as a consequence of the partonic energy loss in the hot and dense QCD medium.

\section{Jet reconstruction techniques}

Before discussing further details of jet quenching measurements at LHC, we 
briefly summarize the jet reconstruction techniques of LHC experiments. All 
three LHC experiments use the \antikt\ jet finding 
algorithm~\cite{aaba:Cacciari} with the distance parameter $R$ varying from 
$0.2-0.6$. ALICE uses tracking plus $\pi^0$s reconstructed in the 
electromagnetic calorimeter for the jet finding~\cite{aahn:ALICE}. ATLAS 
typically uses the calorimeter information only to define 
jets~\cite{baam:ATLAS}. CMS uses the particle-flow algorithm which combines the 
information from the tracking and calorimeter~\cite{aabf:CMS}. All three 
experiments perform the subtraction of the underlying event (UE) present 
underneath the jet reconstructed in heavy ion collisions. The procedure of UE 
subtraction is based on, or similar to, techniques described in 
Refs.~\cite{aabb:Cacciari,aabc:Cacciari,aabd:Cacciari}. A typical transverse 
energy of the underlying event in the 0-10\% most central heavy ion collisions 
per area of $R=0.4$ jet exceeds 100~GeV~\cite{baah:ATLAS,aada:ALICE} which is 
often above the measured jet energy. A careful evaluation of the jet energy 
scale (linearity), jet energy resolution, and jet reconstruction efficiency is 
therefore needed in order to verify the performance of the jet reconstruction 
as stressed by theorists after the asymmetry observation 
measurement~\cite{aabe:Cacciari}.

\begin{figure}
\centerline{
\includegraphics[width=0.9\textwidth,trim=4 4 4 4,clip]{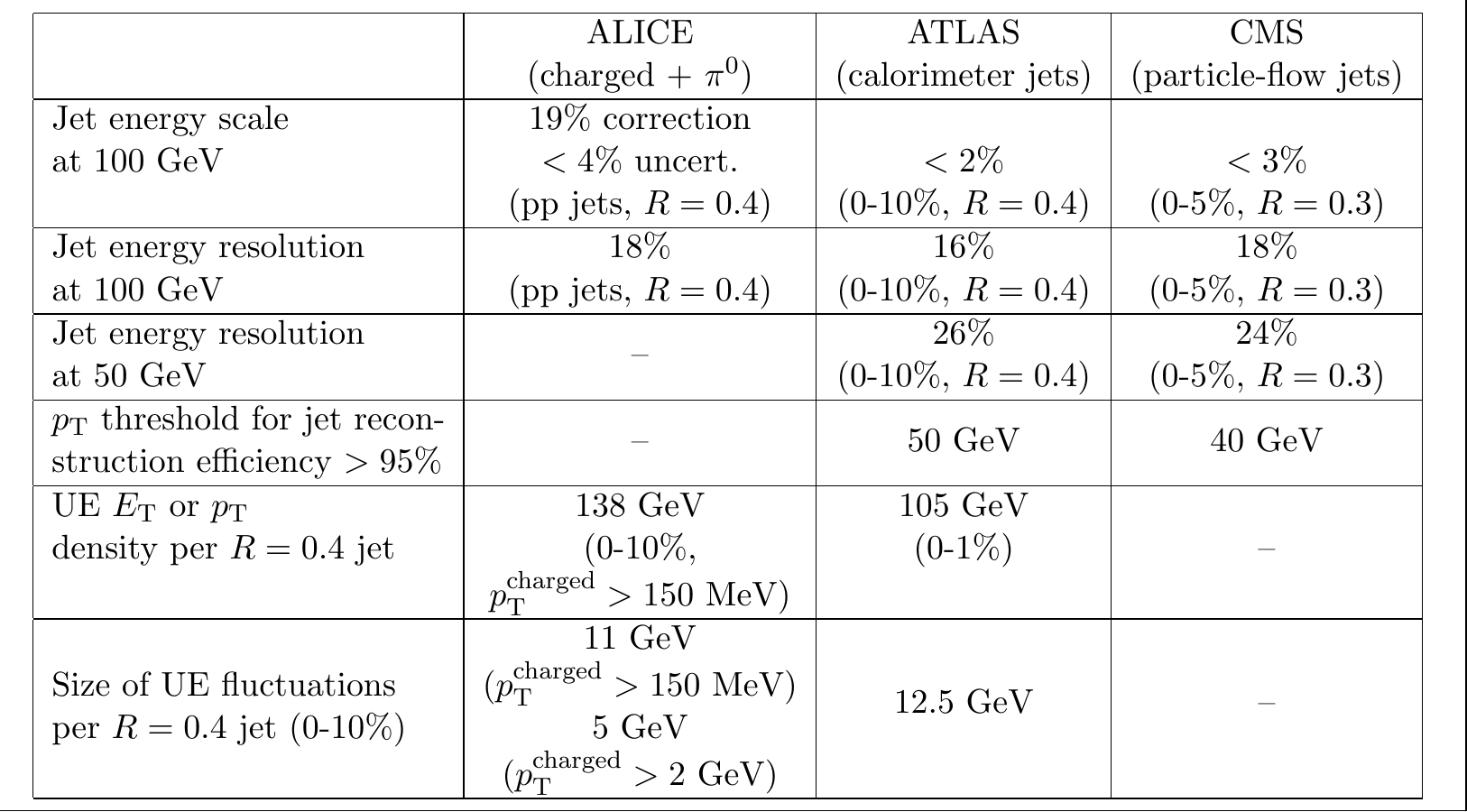}
}
 \caption{
  Benchmarks for the jet reconstruction performance in terms of jet energy 
scale, jet energy resolution and jet reconstruction efficiency. Characteristics 
of UE in terms of mean transverse energy or momentum density of UE per area of 
$R=0.4$ jet or mean size of UE fluctuations are also indicated. The centrality
selection or choice of the jet distance parameter $R$ is indicated in brackets. 
Values are based on plots or explicit numbers given in 
Refs.~\cite{baah:ATLAS,aabq:CMS,aads:CMS,aada:ALICE,aahm:ALICE,aahn:ALICE}.
 }
\label{tbl:perform}
\end{figure}

Several benchmarks summarizing the jet performance in heavy ion events of three 
LHC experiments are summarized in the table on Fig.~\ref{tbl:perform} based on 
Refs.~\cite{baah:ATLAS,aabq:CMS,aads:CMS,aada:ALICE,aahn:ALICE,aahm:ALICE}. Usual techniques to 
determine the jet performance use the embedding of PYTHIA~\cite{aaat:Sjostrand} 
jets into the minimum bias heavy ion background which is either modeled by 
HIJING~\cite{aaav:Wang} and HYDJET~\cite{aaaw:Lokhtin} MC generators or, 
optimally, taken directly from the data. The full detector simulation of these 
events is then performed. 
  The full simulation of embedded events and data-driven tests of jet 
performance discussed further represent a reliable tools to access the impact 
of large UE on the jet reconstruction\footnote{ The conclusions on jet 
performance are reliable up to a possible influence of jet reconstruction by 
strongly modified fragmentation~\cite{aagv:STAR}. We shall see that the 
modifications of fragmentation are moderate and we can therefore expect no 
large impact on performance. It is worth to stress that in-vacuum fragmentation 
is not unique but consists of many configurations of jet properties some of 
which may be close to the jet properties of medium modified jets. Thus, we 
could evaluate the performance as a function of ``fragmentation'' that is e.g. 
as a function of leading particle momentum fraction to make sure that the 
performance of in-medium jets does not suffer from biases.}. Attempts to 
quantify the impact of UE subtraction on jet reconstruction using toy models of 
detector response~\cite{aabe:Cacciari,aadu:Apolinario} are fruitful but not 
always straightforward to connect with specific measurements due to the 
over-simplification of the detector response.

Beside MC evaluation of the jet performance, the data driven tests have been performed.
  Systematic uncertainty due to the jet energy scale and jet reconstruction efficiency 
can be constrained or reduced by matching calorimeter jets to track jets~\cite{aabg:CMS}. The ratio of transverse momentum of jets reconstructed in the 
calorimeter to the transverse momentum of jets reconstructed in the tracking 
system can be compared between data and MC which helps determining systematic 
uncertainties~\cite{baah:ATLAS}.
  Systematic uncertainty due to jet energy resolution can be constrained by 
evaluating fluctuations in the UE in minimum bias events and comparing them to 
fluctuations of energy in reconstructed 
jets~\cite{aada:ALICE,baan:ATLAS,baah:ATLAS}. For calorimeter jets, jet energy 
resolution is given by a direct sum of stochastic and constant term which 
reflect the intrinsic properties of the detector, and by the noise term which 
reflects the fluctuations of UE~\cite{aabu:Fabjan}. This allows for a 
data-driven cross-check of jet energy resolution~\cite{babf:ATLAS}. Similar data 
driven tests have been performed for track jets~\cite{aada:ALICE}. These 
data-driven tests allow to determine systematic uncertainties and help to 
verify the detector response. It is desirable for the experimental community to 
continue investigating such methods since they directly lead to better 
precision of the measurement.

\section{Inclusive jet suppression}
\label{sec:ijs}

The measurement of di-jet asymmetry indicates that observed modifications of 
jets are due to the jet quenching. The di-jet asymmetry is an observation measurement. Alone, it 
cannot provide details on the parton energy loss since it is insensitive to a 
configuration where two jets in a di-jet pair loose comparable amount of energy. 
The measurement of inclusive jet spectra is therefore the first step towards extracting 
detailed features of the jet suppression. 
  The observable quantities of interest are nuclear modification factors, jet 
\Rcp\ and jet \Raa, which characterize the suppression of jet in a given 
centrality interval with respect to peripheral events (\Rcp) 
or with respect to p+p jet spectra (\Raa),

\begin{equation}
\Rcpcent(\pt) = \frac{\ANcollperf}{\ANcollcent} \frac{\hat{N}^\mathrm{cent}(\pt)}{\hat{N}^{60-80}(\pt)},
\Raacent(\pt) = \frac{1}{\ANcollcent} \frac{\hat{N}^\mathrm{cent}(\pt)}{\hat{N}^{\mathrm{pp}}(\pt)},  
\end{equation}
  where $\ANcoll$ is the average number of nucleon-nucleon collisions in a given 
centrality interval~\cite{aaho:Miller} and $\hat{N}(\pt)$ is a per event jet yield. 

\begin{figure}[ht]
\centerline{
\includegraphics[width=0.49\textwidth]{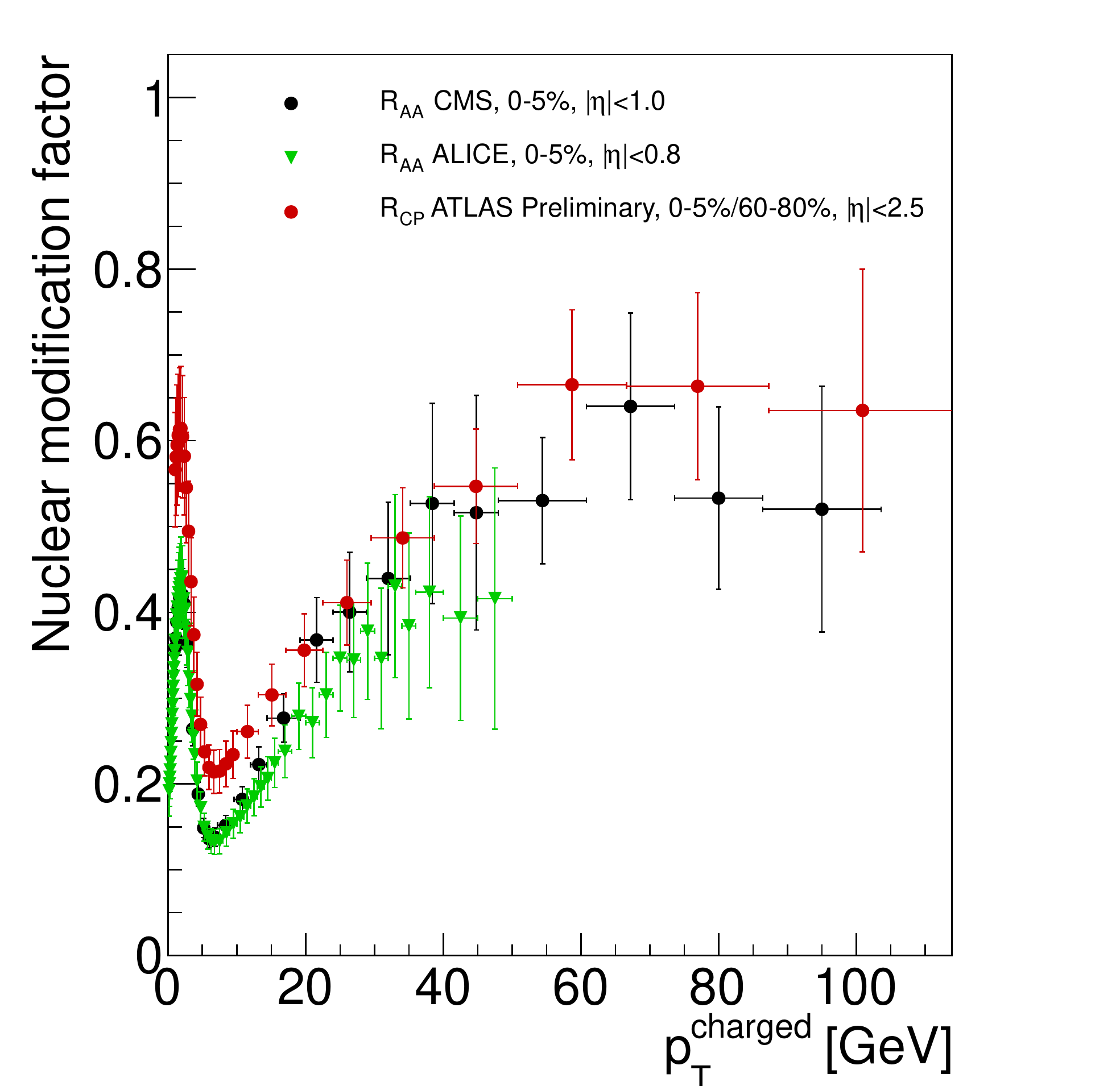}
\includegraphics[width=0.49\textwidth]{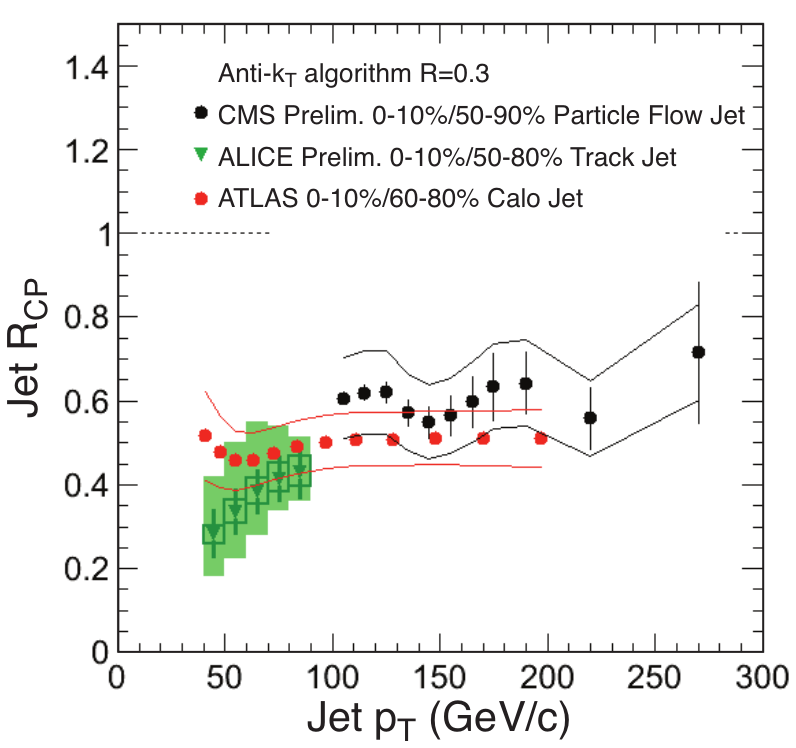}
}
\caption{
  {\it Left:} Nuclear modification factor of charged particles by ALICE~\cite{aadv:ALICE}, ATLAS~\cite{baal:ATLAS}, and CMS~\cite{aabj:CMS}.
{\it Right:} Nuclear modification factor of jets at LHC from Ref.~\cite{aabs:CasalderreySolana}.
  }
\label{fig:Raa}
\end{figure}

The jet \Rcp\ has been measured by ATLAS~\cite{baah:ATLAS} over the \pt\ range 
of $38-210$~GeV. The jet \Raa\ has been measured by CMS~\cite{aabq:CMS} over the 
\pt\ range of $100-300$~GeV and by ALICE~\cite{aahn:ALICE} using jets reconstructed 
from charged particles plus $\pi^0$s over the \pt\ range of $30-120$~GeV. The jet 
yields measured in central collisions are observed to be suppressed by a factor 
of about two at high-\pt. Both ATLAS and CMS see that the suppression is 
consistent with flat \pt-dependence. For $\pt < 100$~GeV ALICE preliminary result
shows a trend of a decrease of \Raa\ with decreasing \pt\ whereas the \Rcp\ 
of ATLAS remains rather flat. A comparison of nuclear modification factors 
measured by different experiments is shown on Fig.~\ref{fig:Raa}.

ATLAS has reported a significant dependence of the inclusive jet suppression on 
the jet size for jets with $\pt < 100$~GeV. For jets with $\pt > 100$~GeV, only 
jets reconstructed with the distance parameter of $R=0.5$ might be different 
compared to jets reconstructed using smaller distance parameters. This is 
consistent with the result by CMS which reports no dependence of the 
suppression on the jet size for $R=0.2-0.4$ jets with $\pt > 100$~GeV. It is 
important to note that any comparison based on jet reconstructed using 
different size parameters might introduce biases since the same jets but 
reconstructed with smaller radii will tend to populate a lower \pt\ 
region~\cite{baah:ATLAS}. E.g. a typical 100 GeV jet 
reconstructed in p+p collisions with $R=0.4$ deposits 15\% of its 
energy in the region of $R > 0.2$~\cite{babd:ATLAS}.
  More details on features seen in the nuclear modification factor of jets is 
discussed in Sec.~\ref{sec:discussion} of this review.

\section{Jet suppression via measurements of single particles}   
\label{sec:sp}

The first observation of jet quenching at RHIC was based on a suppression seen in 
the nuclear modification factor of charged particles~\cite{aaal:STAR,aaak:PHENIX} 
and based on di-hadron azimuthal correlations~\cite{aafb:STAR}. Later, it was 
found that the nuclear modification factor of single charged particles alone is 
insufficient to distinguish different scenarios of microscopical interaction of 
partons with the QCD medium~\cite{aafl:Cole,aafm:Bass}.
  At LHC, the full jet reconstruction with large statistics of jets with high 
transverse momenta is available. Nevertheless, measurements of single particles 
at high-\pt\ is still important for constraining the energy loss models at LHC 
energies~\cite{aahj:Renk}. Moreover, it is crucial for understanding the first 
measurements of jets:
 %
  the nuclear modification factor of charged particles at 
high-\pt\ together with the nuclear modification factor of jets put constraints to 
a suppression of jet fragments at high-\pt\ (or vice versa). If the electroweak processes are 
excluded, each charged particle at high-\pt\ has to be connected with a hard 
scattering and jet production. The nuclear modification factor of inclusive 
charged particles measured by ALICE~\cite{aadv:ALICE}, CMS~\cite{aabj:CMS}, and 
ATLAS~\cite{baal:ATLAS} is shown on Fig.~\ref{fig:Raa}. One can see that the nuclear 
modification factor of charged hadrons at high transverse momentum, $\pt \gtrsim 
40$~GeV, reaches approximately $0.5-0.6$ which are similar values to those 
measured in the nuclear modification factor of jets discussed in 
Sec.~\ref{sec:ijs}. This implies that there should be no substantial modification 
of jet fragmentation at high-\pt. This was indeed observed in the jet 
fragmentation measurement which is discussed in Sec.~\ref{sec:ff}. Three 
independent measurements -- nuclear modification factor of single hadrons, nuclear 
modification factor of jets, and jet fragmentation functions -- are therefore in a 
good agreement with each other and also in a good agreement among experiments.

Below $\pt \approx 20$~GeV the nuclear modification factor exhibits similar 
behavior as seen at RHIC~\cite{aaap:BRAHMS,aaam:PHENIX,aaao:PHOBOS,aaan:STAR}. The 
minimum of \Raa\ at LHC is reached at $\pt = 6-7$~GeV with the value of $0.13-0.14$ 
which is about 50\% less than at RHIC. The rise of the \Raa\ for \pt\ above 8 GeV 
is predicted by most of the jet quenching models~\cite{aafk:Abreu}, however the 
magnitude of the predicted slope varies greatly among models. For a comparison 
of \Raa\ with different jet quenching models see e.g. Refs.~\cite{aahl:Renk,aadv:ALICE,aabj:CMS}.

The consistency of results on suppression of inclusive jet yields is further checked by the 
measurement of particle-yield modifications of jet-like di-hadron azimuthal
correlations done by ALICE~\cite{aago:ALICE} and CMS~\cite{aabr:CMS}. The 
modifications of di-hadron azimuthal correlations is quantified by the \Iaa\ 
modification factor defined as a ratio of integrated per-trigger-particle 
associated yield measured in Pb+Pb collisions to the yield measured in p+p 
collision. 
 %
  Both ALICE and CMS report a suppression of the away-side yield by a factor of 
$0.5-0.6$ in central Pb+Pb collisions which does not show any strong \pt\ 
dependence. This is again consistent with the inclusive jet suppression 
discussed in Sec.~\ref{sec:ijs}. Another interesting feature of the data on 
\Iaa\ is the fact that the near-side yield is not suppressed. At first glance, this is 
suggestive of a ``corona jet production''~\cite{aabr:CMS} that is 
the predominant jet production at the surface of the medium and strong 
absorption in the inner core~\cite{aagr:Pantuev}.
  No suppression of nearside jet yield might also be explained as 
a consequence of non-trivial interplay between the modification of jet 
fragmentation and different suppression of quark and gluon jets (discussed 
further in Sec.~\ref{sec:ff}) both of which can lead to a bias in the selection 
of the trigger particle in Pb+Pb collisions~\cite{aahf:Renk}. 

\begin{figure}
\centerline{
\includegraphics[width=0.95\textwidth]{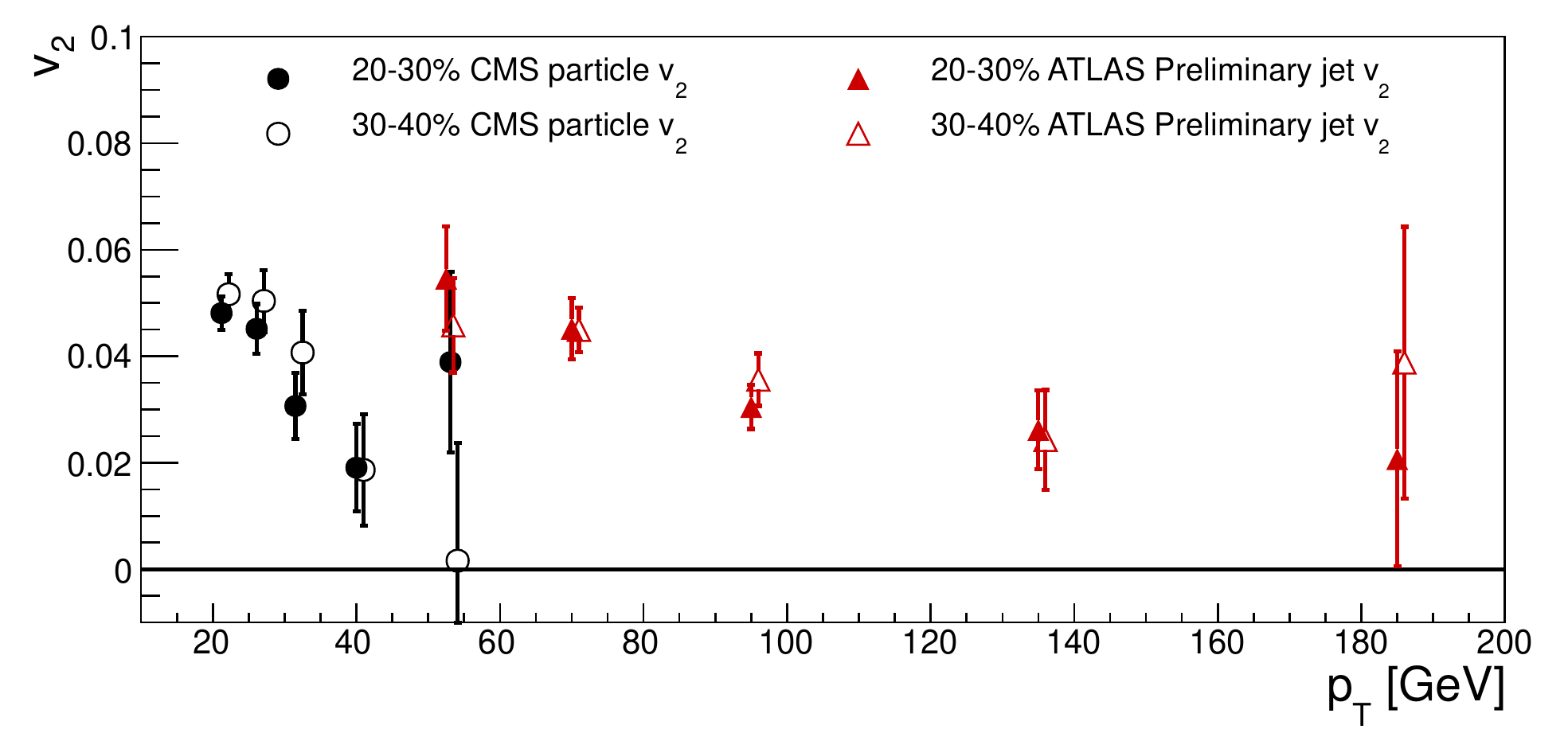}
}
\caption{
  Comparison of $v_2$ at high-\pt\ for two centrality bins. Results on charged 
particle $v_2$ at high-\pt\ of CMS~\cite{aabl:CMS} are shown along with results 
on $v_2$ of jets by ATLAS~\cite{baau:ATLAS}.
  }
\label{fig:azim}
\end{figure}

\section{Azimuthal dependence of the jet suppression}

The measurement of azimuthal dependence of charged particle production at 
high-\pt\ along with the measurement of nuclear modification factor can help 
distinguishing the microscopic mechanisms of parton energy loss through 
determining the path length ($l$) dependence of the energy loss, $\Delta E \approx 
l^{\alpha}$ 
~\cite{aafu:Baier,aafs:Peigne,aafo:Gubser,aacr:Wicks,aaga:Marquet,aafv:Renk,aafy:Betz,aafw:Jia}.
   Measurement of azimuthal anisotropy of $\pi^0$ with $\pt=7-10$ GeV at RHIC 
supported a scenario of $\alpha = 3$ based on AdS/CFT gravity-gauge dual 
modeling~\cite{aafp:Liu,aagb:Gubser,aafo:Gubser,aaga:Marquet}, while solely 
perturbative QCD
  calculations 
  ($\alpha = 1$ for collisional energy loss and 
$\alpha = 2$ for radiative energy loss) seemed to under-predict the 
data~\cite{aacg:PHENIX}. 
   Recently, it has been shown that it may be problematic to distinguish the 
weakly coupled pQCD-based medium and strongly coupled AdS/CFT-modeled medium
   based only on the path length dependence of light quark jets~\cite{aacd:Betz}.
   Nevertheless, the measurement of the path-length dependence 
remains essential for understanding the parton energy loss.
   To get more insight into the path-length dependence of the jet suppression 
the azimuthal anisotropy of high-\pt\ particles is measured by 
ALICE~\cite{aagm:ALICE}, ATLAS~\cite{baae:ATLAS}, and CMS~\cite{aabl:CMS}. 
Further, the azimuthal anisotropy of jets is measured by 
ATLAS~\cite{baau:ATLAS}.

   The azimuthal anisotropy is characterized by the second-order Fourier 
harmonic coefficient ($v_2$) in azimuthal angle ($\phi$) which is a measure of 
hadron emission relative to the reaction plane angle 
($\Psi_{\mathrm{RP}}$)~\cite{aagn:ALICE,baae:ATLAS,aaft:CMS}, 

\begin{equation}
\frac{\mathrm{d} N}{\mathrm{d} (\phi - \Psi_{\mathrm{RP}})} \approx 1 + 2v_2 \cos(2(\phi- \Psi_{\mathrm{RP}} )).
\end{equation}
  Such azimuthal anisotropy at low 
transverse momenta ensues because the hadron yield is more suppressed along the 
long axis of the almond shaped fireball than the short axis. Thus the magnitude 
of the $v_2$ is sensitive to the path length dependence of 
energy loss~\cite{aaaj:Wang}. The measured $v_2$ of charged particles increases 
at low \pt\ achieving a maximum at $\pt \approx 3$ GeV and then gradually 
decreases towards zero but remains positive to at least 40 GeV.
  For charged particles with $\pt$ in the interval of $14-48$~GeV the $v_2$ 
lies approximately between 0.02 and 0.08. Even in the high-\pt\ region, the 
$v_2$ has a characteristic centrality dependence with $v_2$ being generally 
larger in less central collisions where the eccentricity of initial overlap 
region is larger compared to the most central collisions. This result is 
consistent among all there experiments though CMS provides the measurement up 
to $\pt \approx 60$~GeV, whereas ALICE and ATLAS only up to $\pt=20$~GeV.

The results on the $v_2$ of high-\pt\ charged particles are qualitatively 
consistent with results on jet $v_2$ by ATLAS which measures the jet $v_2$ for 
$\pt = 50-210$~GeV. The non-zero jet $v_2$ is distinguishable up to $\pt=160$~GeV. The 
variation of the jet yield with the distance to the reaction plane, $\Delta 
\phi$, is also characterized by the ratio of jet yields between the most 
out-of-plane ($\frac{3\pi}{8} < \Delta \phi < \frac{\pi}{2}$) and most in-plane 
($0 < \Delta \phi < \frac{\pi}{8}$) jet directions. This ratio shows as much as 
20\% variation between the out-of-plane and in-plane jet yields. A comparison 
of measurements is presented on Fig.\ref{fig:azim}.



\section{Photon-jet and $Z^0$-jet correlations}

  The fact that the measured jet suppression at LHC is not an effect of initial 
state modifications was proven by the measurement of inclusive yields of prompt 
photons ($\gamma$)~\cite{baar:ATLAS,aabi:CMS} and inclusive yields of $W^\pm$, 
$Z^0$ vector bosons~\cite{baai:ATLAS,baak:ATLAS,aabh:CMS} which do not exhibit 
any suppression in central heavy ion collisions. Not only no suppression as a 
function of \pt\ is seen, but also no modification of the rapidity distribution 
of produced vector bosons is observed which is also sensitive to modifications 
of nuclear parton distribution functions~\cite{aafn:Paukkunen}. 

  Since photons and vector bosons are insensitive to the colored medium they 
can provide, on a statistical average, constraints on the energy of the 
away-side parton shower~\cite{aadp:Wang}.
  The $\gamma$-jet and $Z^0$-jet events has indeed been studied by 
ATLAS~\cite{babb:ATLAS,baay:ATLAS} and CMS~\cite{aabo:CMS} and the energy loss has 
been evaluated in a similar way as for the original asymmetry 
measurement~\cite{baaa:ATLAS,aaab:CMS}. Instead of the asymmetry, a more simple 
quantity has been evaluated -- the mean fractional energy carried by a jet, 
$x_{\mathrm{J}\gamma} = \ptjet/\ptgamma$ or $x_{\mathrm{J}Z^{0}} = \ptjet/\ptZ$. 
The conclusion from these measurements was the same as from the original 
measurement of the di-jet asymmetry: a significant increase of events with large 
imbalance of transverse momentum in the $\gamma$-jet or $Z^{0}$-jet system has been 
observed, while the $\Delta \phi$ distribution remains with no modification.

Measurements of the jet suppression in the $\gamma$-jet or $Z^{0}$-jet system 
provide a starting point for detailed quantification of energy loss. 
   In principal, by selecting a suitable range for the transverse momentum of 
the tagging photon accessible at different center-of-mass energies, the 
in-medium modification of parton showers in the dense matter created at these 
different center-of-mass energies can be studied.
   First theoretical study 
towards this goal were done in Ref.~\cite{aado:Dai}. 
   Besides tagging the energy of a quenched jet by the energy of $\gamma$ or 
$Z^0$, the $\gamma$-jet or $Z^0$-jet system can provide a tool to study the 
difference between quark and gluon energy loss. This is in principle possible 
because the quark-to-gluon fraction of partons initiating the jets in these 
systems is on average different from the quark-to-gluon fraction of di-jets.
   $Z^0$-jet and $\gamma$-jet systems are also convenient from the experimental 
point of view since $Z^0$ or $\gamma$ is less sensitive to resolution effects 
comparing to jets and therefore the unfolding of observables that depend on the 
kinematics of both, the jet and the vector boson, is less demanding than 
the unfolding of the di-jet observables.
  The clear disadvantage of these measurements are low yields due to smallness 
of electromagnetic coupling. Nevertheless, the $Z^0$-jet and $\gamma$-jet 
measurements represent one of the golden channels for a long term heavy-ion 
program after the upgrade of LHC~\cite{babe:ATLAS,aafa:CMS}.

\section{Flavor dependence of the jet suppression}
\label{sec:bjet}

The quenching of jets in heavy-ion collisions is expected to depend on the flavor 
of the fragmenting parton. Jets initiated by heavy quarks are expected to radiate 
less than jets initiated by light quarks. The gluon radiation of a heavy quark is 
suppressed at angles smaller than the ratio of the heavy quark mass to its energy. 
This is so-called dead-cone effect~\cite{aacj:Dokshitzer}. Consequently, the heavy 
flavor jets are expected to undergo a smaller suppression than the light quark 
jets. However, measurements of heavy quark production at RHIC via semi-leptonic 
decays to electrons showed a combined charm and bottom suppression in Au+Au 
collisions comparable to that observed for inclusive hadron 
production~\cite{aack:PHENIX,aacl:STAR,aacm:PHENIX}. There is disagreement in the 
theoretical literature regarding the interpretation of the RHIC heavy quark 
suppression measurements~\cite{aacr:Wicks,aacn:Djordjevic,aaco:Gossiaux,aacp:Uphoff} particularly regarding 
the role of non-perturbative effects~\cite{aadm:Horowitz,aadl:VanHees,aadk:Adil}.

    To access the difference between the jet suppression of heavy quarks and light 
quarks CMS measures the $b$-tagged jets~\cite{aabt:CMS}, ALICE and ATLAS measure the 
semi-leptonic decays of open heavy flavor hadrons into muons~\cite{baap:ATLAS,aagp:ALICE}. 
The measurement of $b$-jets by CMS uses the secondary vertex mass distribution to 
tag the jets initiated by a bottom quark. The main quantity extracted from the 
measurement is the bottom quark jet to inclusive jet ratio. This $b$-jet fraction is 
measured in the range of $80 < \ptjet < 200$ GeV both in Pb+Pb and p+p data 
at $\sqrtsnn = 2.76$~TeV. The extracted $b$-jet fraction lies in the range of 
$2.9-3.5\%$ 
  and is comparable between Pb+Pb and p+p on one side and Pb+Pb and MC on the 
other side. The $b$-jet fraction as a function of \ptjet\ does not exhibit any 
centrality dependence though the systematic uncertainty is rather large. The 
measured $b$-jet fraction is compatible with approximately half a percent 
decrease with increasing \ptjet\ predicted by PYTHIA for the \ptjet\ range 
covered by the measurement. These observations imply that the $b$-jet \Raa\ 
(which can be defined as a product of $b$-jet fraction in Pb+Pb and p+p) and the 
inclusive jet \Raa\ exhibits the similar behavior as the inclusive jet \Raa: 
suppression by a factor of two, possibly only a modest dependence on \ptjet. 
Although, it needs to be stressed that the systematic and statistical 
uncertainties on that measurement are large.

  The measurement of semi-leptonic decays of open heavy flavor hadrons by 
ATLAS~\cite{baap:ATLAS} is performed in the midrapidity ($|\eta|<1.05$) and over 
the muon transverse momentum range $4 < \pt < 14$~GeV. The measurement by 
ALICE~\cite{aagp:ALICE} is performed in forward rapidity ($2.5<|\eta|<4$) and 
over the muon transverse momentum range $4 < \pt < 10$~GeV. Over these \pt\ 
ranges, muon production results predominantly from a combination of charm and 
bottom quark semi-leptonic decays. ATLAS compares the differential yields of 
muons in a given centrality and \pt\ bin to yields measured in peripheral 
collisions via central-to-peripheral ratio, \Rcp, whereas ALICE evaluates the 
difference in terms of \Raa\ that is with respect to the p+p reference measurement. 
Both experiments see a smooth decrease of nuclear modification factor with 
increasing centrality. The suppression does not change with the muon \pt\ while 
the size of the suppression changes by nearly a factor of two between the most 
central and the most peripheral collisions (ATLAS), resp. by a factor of $3-4$ 
between the yields measured in the most central Pb+Pb collisions and yields 
measured in p+p (ALICE).
  The results of ATLAS measured in the midrapidity region can be also compared with 
results of CMS measurement of high-\pt\ non-prompt $J/\psi$ ($\pt = 
6.5-30$~GeV) produced in the decay of $b$-hadrons~\cite{aagt:CMS}. The result of 
CMS evaluated in terms of \Raa\ in the 0-20\% collisions are consistent with 
the \Rcp\ of ATLAS. However, opposed to ATLAS, CMS does not report any 
centrality dependence of \Raa\ of non-prompt $J/\psi$;
  the result in 20-100\% centrality bin is consistent with the result in central 0-20

  Both measurements, the $b$-jet fraction and semi-leptonic decays of open 
heavy flavor seem to suggest that there is no dramatic difference between the 
suppression of inclusive jets and jets initiated by $b$-quarks. Although, as 
already stated above, more precision is needed on the direct $b$-jet measurement 
to make this conclusion stronger.


\section{Jet internal structure}
\label{sec:ff}

One of the tools that allow precise comparison between the data and theoretical 
models of jet quenching is the measurement of jet fragmentation. The key question 
to address is how are the parton showers modified by the medium. Generally, there 
is no agreement in the theory upon the answer to this question, however most of 
the theoretical models predict a softening of fragmentation functions -- 
suppression at large momenta of fragments, enhancement at small momenta of 
fragments, and broadening of a jet~\cite{aadh:Salgado,aadc:Lokhtin,aadw:Borghini,aacu:Armesto,aadd:Vitev,aady:Armesto,aadz:Renk}.
 %
 %
 %
 %
    The softening of fragmentation was first indirectly observed by CMS in the 
measurement of the missing-\pt~\cite{aaab:CMS}. In that measurement,  
the projection of missing-\pt\ of reconstructed charged particles onto 
the leading jet axis was calculated as $\mpt = \Sigma_i \pti \cos(\phi_i - 
\phi_{\mathrm{leading~jet}})$ for each event. The event averaged missing-\pt , $\langle \mpt \rangle$, was 
then evaluated as a function of the di-jet asymmetry in three configurations -- 
inside the leading and subleading jet cone, outside the leading and subleading 
jet cones, and inclusively. An in-cone imbalance of $\langle \mpt \rangle 
\approx 20$~GeV was found for di-jet events with the largest asymmetry both in 
MC and data. This in-cone imbalance is balanced by the out-of-cone imbalance of 
$\langle \mpt \rangle \approx 20$ GeV again both for MC and data. However, in 
the data the out-of-cone contribution is carried almost entirely by tracks with 
$0.5 < \pt < 4$~GeV whereas in MC more than 50\% of the balance is carried by 
tracks with $\pt > 4$~GeV, with a negligible contribution from $\pt < 1$~GeV. 
The MC events with large di-jet asymmetry are due to the semi-hard initial or 
final state radiation. This implies that the missing-\pt\ in those MC events is 
balanced by rather hard-\pt\ particles which are very often clustered into a 
third jet. On the contrary, the results for large di-jet asymmetry in the data show 
that a large part of the momentum balance is carried by soft particles radiated 
at large angles to the jet axes. 
   Another important feature seen in the measurement is that the overall 
momentum balance in the event is recovered when calculating the missing-\pt\ 
using all particles with $\pt>0.5$~GeV. This confirms that the large di-jet 
asymmetry is not due to instrumental effects or e.g. due to neutrino 
production\footnote{There is no a~priory reason why the balance of the event 
should occur for the threshold of $\pt=0.5$~GeV. It would clearly be of 
interest for the theory to extract this threshold with higher precision, if 
experimentally feasible}. The physics picture obtained from missing-\pt\ 
measurement was further confirmed by the measurement of jet-like di-hadron 
correlations by ALICE~\cite{aago:ALICE} and CMS~\cite{aabr:CMS} which we have 
also discussed in Sec.~\ref{sec:sp}. In those measurements, the yield of 
trigger-associated particles is observed to be enhanced in the soft-\pt\ region 
compared to the p+p reference.

\begin{figure}
\centerline{
\includegraphics[width=0.49\textwidth]{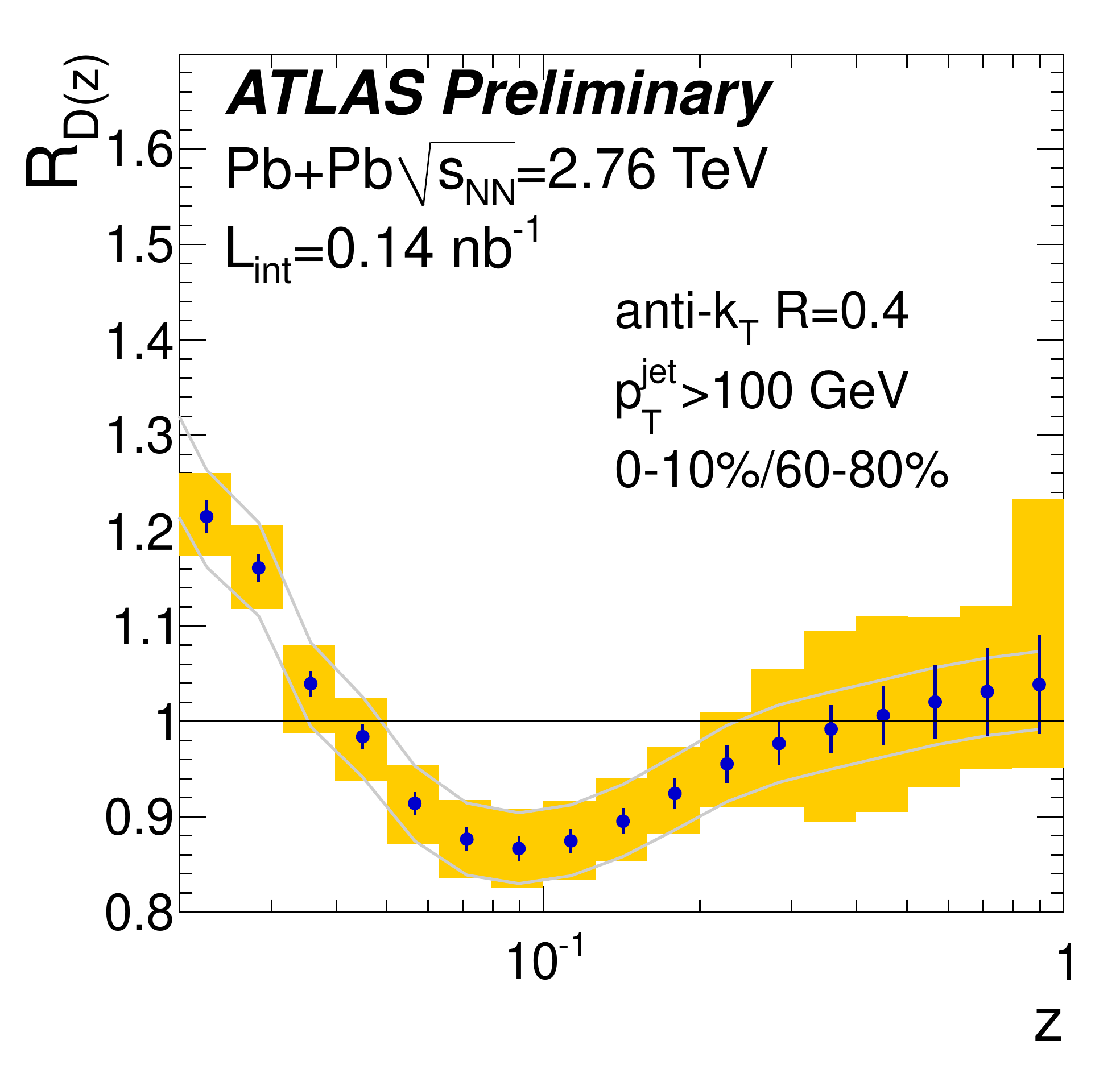}
\includegraphics[width=0.49\textwidth]{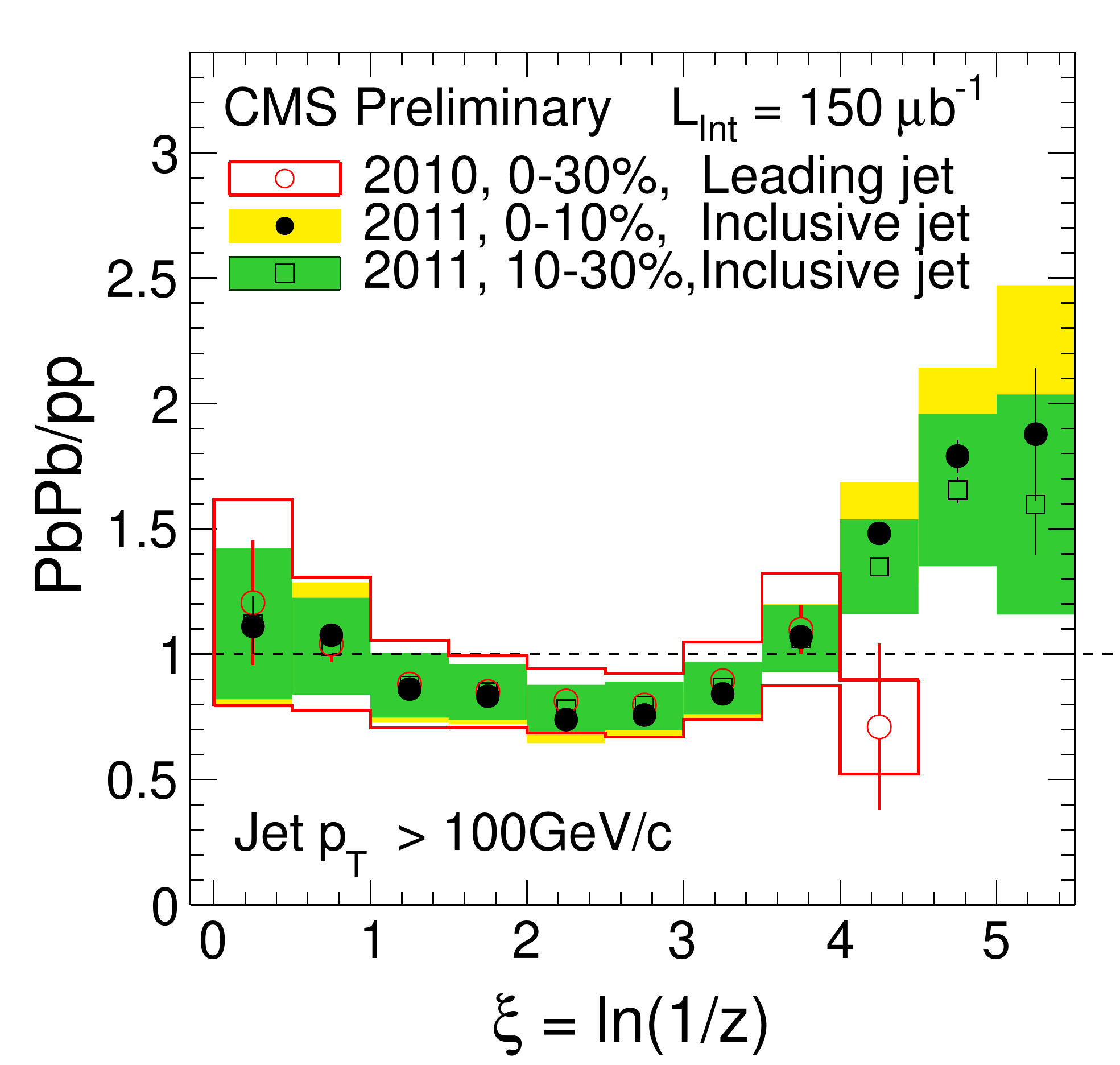}
}
\centerline{
\includegraphics[width=0.49\textwidth]{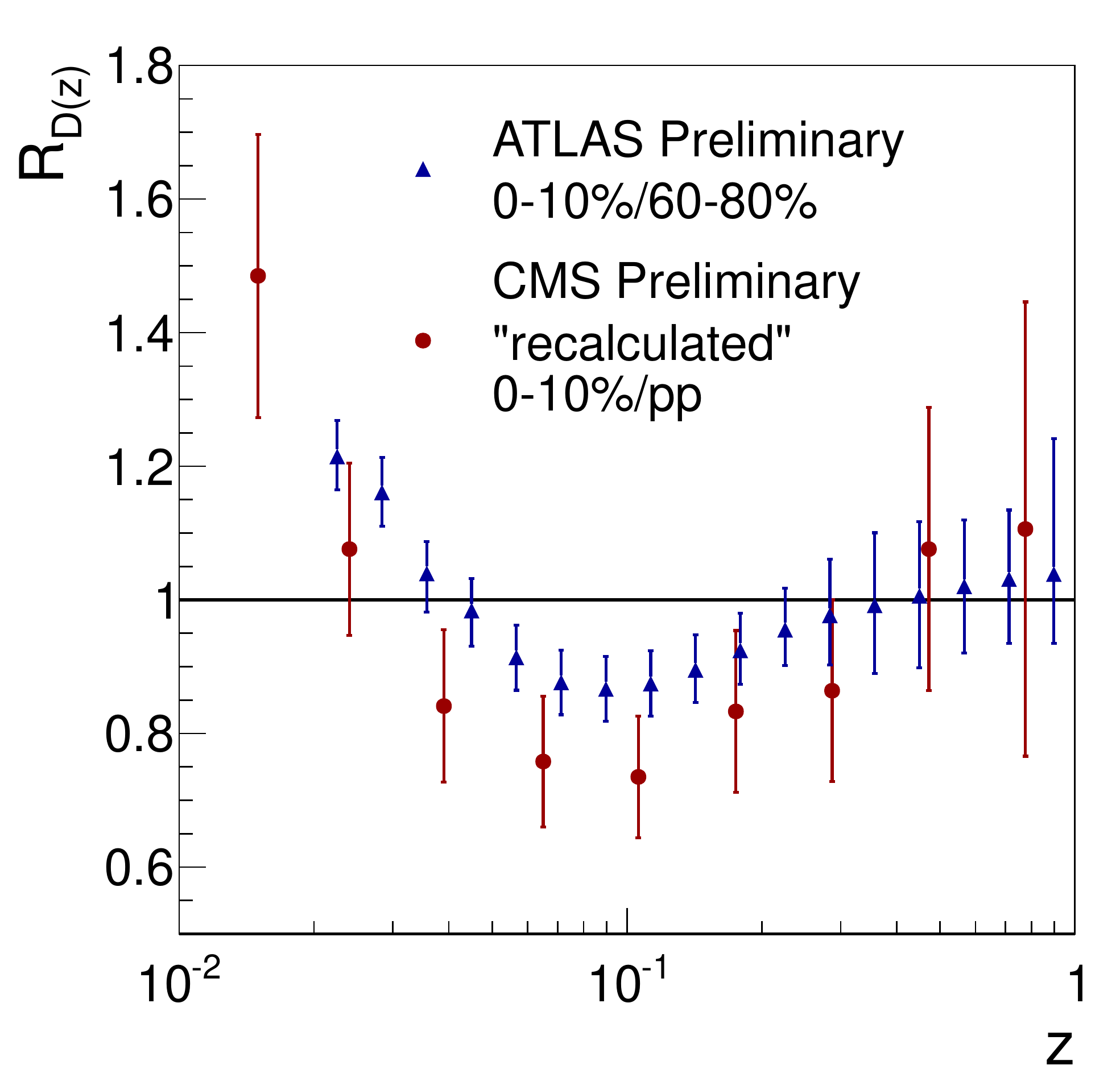}
\includegraphics[width=0.49\textwidth]{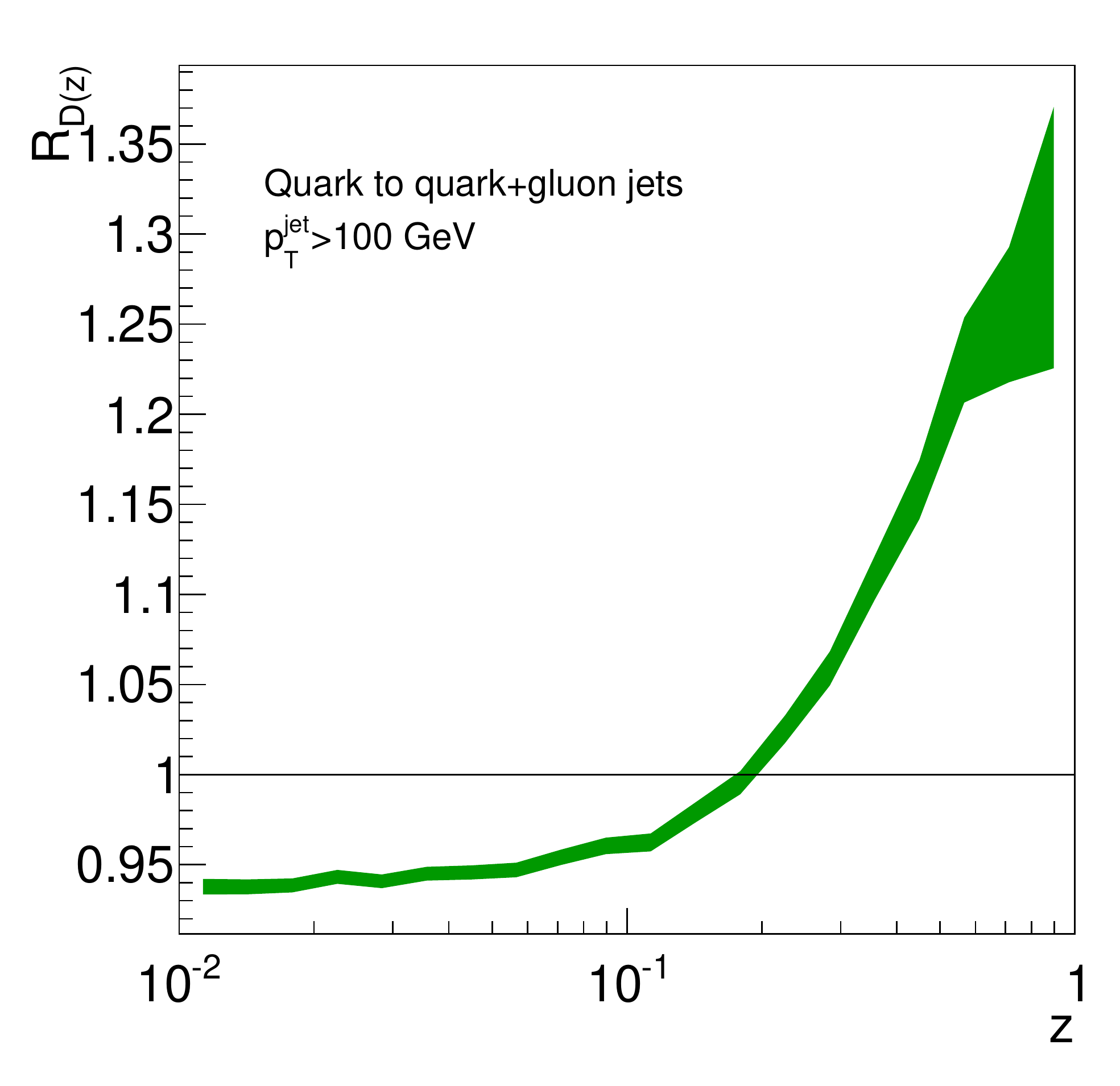}
}
\caption{
  {\it Upper left:} Central to peripheral ratio of fragmentation functions 
measured by ATLAS~\cite{baat:ATLAS}. Fragmentation functions are evaluated in 
terms of momentum fraction, $z$. {\it Upper right:} Ratio of fragmentation 
functions in central Pb+Pb collisions to p+p collisions measured by 
CMS~\cite{aabp:CMS}. Fragmentation functions are evaluated in terms of $\xi \approx
\ln(1/z)$. {\it Lower left:} CMS result recalculated in terms of $z$ and 
compared to the result by ATLAS. Error bars represent combined statistical and 
systematic uncertainties. For differences between ATLAS and CMS measurements 
see the text. {\it Lower right:} Ratio of fragmentation functions in light-quark 
($u,d,s$) jets to inclusive jets (light-quark + gluon jets). Jets 
with $\pt>100$ GeV are considered. Simulation of p+p collisions at 
$\sqrtsnn = 2.76$~TeV by PYTHIA 6.4.
  }
\label{fig:RDz}
\end{figure}

A natural question arises -- what \pt\ region of jet fragments feeds up the 
soft region? This question can be addressed by a direct measurement of 
fragmentation functions. The first measurements of the fragmentation 
functions~\cite{aabn:CMS,baam:ATLAS} were consistent with no-modification due to 
the large systematic and statistical uncertainties. However, the latest 
preliminary results by ATLAS~\cite{baat:ATLAS} and CMS~\cite{aabp:CMS} show a 
significant change in the structure of fragmentation functions between central 
and peripheral collisions. Both experiments evaluate the fragmentation for jets 
having the same minimum-\pt\ threshold of 100~GeV, although otherwise there are 
important differences between these two measurements:

  \begin{itemize}
  \item ATLAS uses radial distance of $\Delta R=0.4$ to match particles to jets 
that were reconstructed with different radii ($R=0.2,0.3,0.4$); CMS uses $\Delta R=0.3$ for jets 
reconstructed with the radius of $R=0.3$.
  \item ATLAS evaluates the fragmentation functions in terms of variable $z$, 
$z = (\ptparticle/\ptjet) \cos \Delta R$ while CMS uses variable $\xi$, 
$\xi = \ln(\ptjet / \ptparticle) \approx \ln(1/z)$. 
  \item ATLAS uses jets in the pseudorapidity interval of $|\eta|<2.1$ while CMS uses interval of $0.3<|\eta|<2$.
  \item ATLAS uses minimum \pt\ threshold for charged particles of 2 GeV while CMS 
uses 1~GeV.
  \end{itemize}

Despite to these differences the two experiments provide qualitatively the same 
result. Both experiments evaluate the modification of fragmentation in terms of 
the ratio of fragmentation functions in central collisions to the fragmentation 
functions in peripheral (ATLAS) or p+p collisions (CMS). In the case of no 
modification this ratio should be unity. The results for the most central 
collisions are shown on upper plots of Fig.~\ref{fig:RDz}. The lower left plot 
of Fig.~\ref{fig:RDz} compares the result by ATLAS with the $\xi$-to-$z$ 
recalculated result by CMS. The recalculated CMS result obviously has to be 
taken as an approximate, but one can directly see a good qualitative agreement 
in these two measurements.
  The enhanced yield of low-$z$ / low-\pt\ / high-$\xi$ particles in central 
collisions is accompanied by a reduction in the yield of intermediate-$z$ / 
$\xi$ / \pt\ particles\footnote{It is worth to note that the estimate of the 
magnitude of low-\pt\ enhancement might be biased throughout the 
necessary subtraction of the UE which might also be modified by the enhancement 
of soft particles as seen in the missing-\pt\ study. The sensitivity to these 
effects might be tested by varying the distance between the jet and the area 
used for the determination of the UE.}.
  No reduction of yields at high-$z$ is seen. On the contrary, rather an 
indication of enhancement can be seen. This may be puzzling in a view of the 
theory that commonly predicted a reduction at 
high $z$~\cite{aadc:Lokhtin,aacu:Armesto,aadw:Borghini,aady:Armesto,aadz:Renk}.
  It is important to note that experiments evaluate naturally the 
fragmentation functions with respect to the jet energy which is different from 
the energy of the original parton due to the quenching. On the other hand, 
theoretical studies commonly evaluate the fragmentation functions with respect to the 
energy of parton prior to the quenching. This makes any quantitative comparison 
between measured data and predicted modifications of fragmentation functions 
difficult.

  Recently, the observed results of modified fragmentation were rather successfully 
reproduced by theoretical calculations~\cite{aadq:Zapp,aadr:Kharzeev}. Both 
references use the quenched jet energy to define the fragmentation functions. Reference~\cite{aadq:Zapp} uses a full simulation of events based on PYTHIA+JEWEL MC 
generators. In this framework, PYTHIA is used to generate the hard-scattering and 
virtuality ordered parton showers. Then, JEWEL generates the final state parton 
shower modified by the medium. Finally, PYTHIA takes over and does the 
hadronization.
  In the reference~\cite{aadr:Kharzeev}, authors use an effective 1+1 dimensional 
quasi-Abelian model to describe non-perturbatively the dynamics of fragmentation of 
quark jets in the medium. Both models are successful in describing the trends seen in 
experimental data even though being based on very different modeling of the jet 
quenching.


\section{Discussion}
\label{sec:discussion}

\begin{figure}
\centerline{
\includegraphics[width=0.49\textwidth]{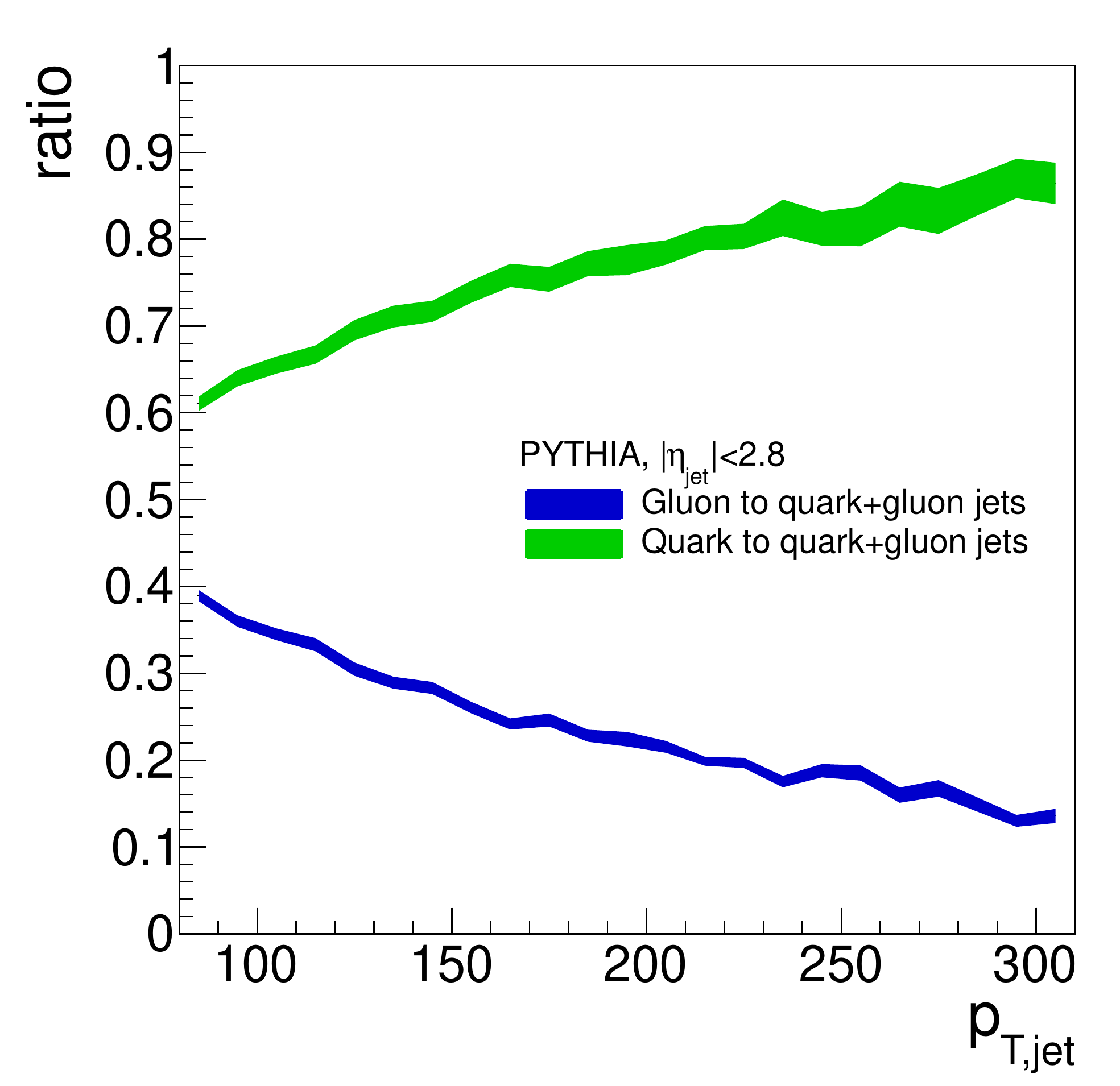}
\includegraphics[width=0.49\textwidth]{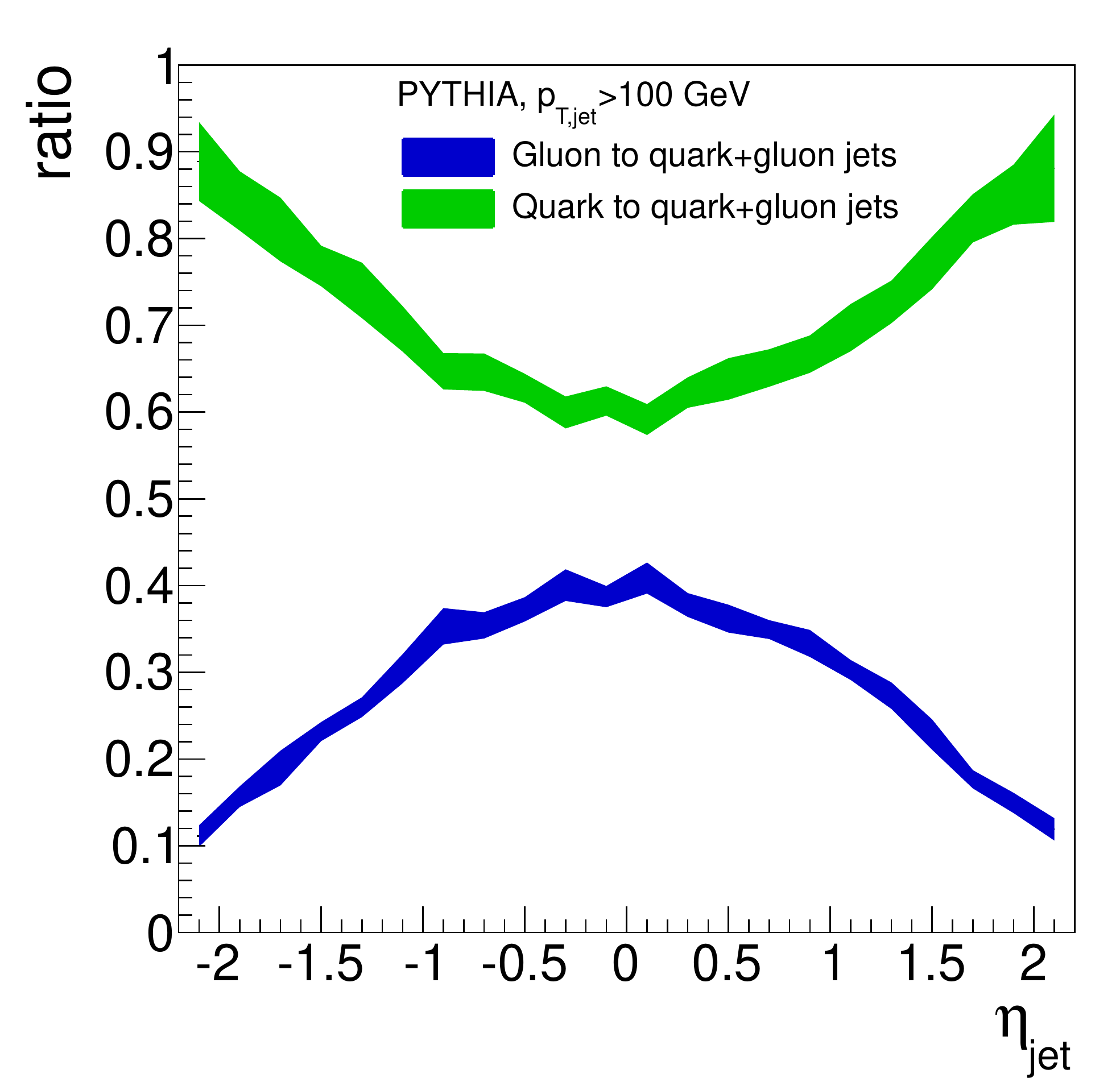}
}
\caption{
  Ratio of multiplicities of jets initiated by gluon (blue) or light-quark 
(green) to multiplicities of inclusive jets (light-quark + gluon 
jets). Ratio is evaluated as a function of jet transverse momentum {\it (left)} 
and jet pseudorapidity {\it (right)}. 
Simulation of p+p collisions at $\sqrtsnn = 2.76$~TeV by PYTHIA 6.4. 
}
\label{fig:QtoG}
\end{figure}

Throughout previous sections we could see a multitude of experimental 
observations of features connected with the jet quenching. 
  Let us first discuss more closely the features seen in the fragmentation 
measurement. Without considering any details on the jet quenching models, it is 
remarkable that the shape of the central-to-peripheral ratio of fragmentation 
functions resembles some features of the ratio of quark jet fragmentation to 
quark+gluon jet fragmentation in the vacuum. This ratio modeled by PYTHIA is 
shown on the lower right plot of Fig.~\ref{fig:RDz}. The characteristic 
difference between the quark and gluon jet fragmentation in the vacuum was 
measured in details at LEP~\cite{aaei:ALEPH,aaej:OPAL}. The difference reflects 
the fact that quark jets are more collimated and contain more high-$z$ 
fragments whereas gluon jets are generally broader and 
softer~\cite{aaef:DeFlorian,aaeg:CDF,aaeh:ALEPH}.
   In p+p collisions at the center-of-mass energy of 2.76~TeV, the quark jets 
should constitute about two thirds of all jets at 100 GeV whereas gluon jets 
should constitute about a third of all jets. This can be seen from the PYTHIA 
estimate of quark-initiated vs. gluon-initiated jet fractions in di-jet events 
shown on Fig.~\ref{fig:QtoG}.
   In the radiative energy loss scheme, gluon jets should be quenched more 
strongly than light quark jets due to the larger color factor for gluon 
emission from gluons than from quarks.
   Thus, the quark-to-gluon jet fraction has to be larger in central compared to 
peripheral collisions.
   It is therefore inevitable that any change in the pattern of fragmentation 
functions in central compared to peripheral collisions (or compared to p+p 
reference) will be modulated by a difference from different quark-to-gluon jet fractions. 
  The ratio of fragmentation functions will be given by a convolution of 
modification of fragmentation due enhanced quark-to-gluon fraction and a 
modification of fragmentation due to the energy loss itself. If we convolute 
the original predictions on medium modified fragmentation
  (predicting generally a significant depletion of 
fragments at high-$z$ and an increase of fragments at low-$z$) with the 
characteristic pattern of fragmentation due enhanced quark-to-gluon fraction 
(implying increase of fragments at high-$z$ and decrease of fragments at low-$z$) 
we might get a pattern measured by ATLAS and CMS.

   The argument of influence of measured fragmentation by the difference in quark-to-gluon jet 
fractions may also provide a simple explanation for an enhanced yield of fragments 
at high-\pt\ that was seen in the measurement by 
  ATLAS\footnote{Note that ATLAS evaluates the ratio of \pt\ distribution 
  whereas CMS evaluates the difference. The difference of \pt\ distribution is not 
sensitive to any modifications at high-\pt\ due to small values of the 
distribution at high-\pt\ (the order of $10^{-2} - 10^{-3}$). The difference might 
also have a different position of minimum than the ratio.}\,:
   the decrease in the yields of fragments at high-\pt\ can be 
compensated by an effective increase due to different quark-to-gluon jet ratio in 
central compared to peripheral collisions.
   Further, the quark-to-gluon jet fractions differ at different 
pseudorapidities as illustrated in the right plot of Fig.~\ref{fig:QtoG}. This 
implies that any measurement performed at different pseudorapidities will 
differ. This might be a reason for the difference between the overall magnitude 
of $R_D(z)$ of ATLAS and CMS seen in the lower left plot of 
Fig.~\ref{fig:RDz}\footnote{It is also worth to stress that the ratio of 
fragmentation functions is sensitive to the overall normalization. 
Fragmentation functions should be normalized by the total number of all jets 
even if they do not have any associated particle with \pt\ above a certain 
threshold (2 GeV in case of ATLAS or 1 GeV in case of CMS). Normalizing only by 
jets that have particles with \pt\ above the threshold leads to a bias since 
there is a larger probability for a soft jet to exceed the threshold in central 
compared to peripheral collisions due to the presence of UE. Even 
if the UE subtraction is correctly performed, the bias in the normalization is 
not removed.}.
   It is clear that these arguments are qualitative, based on the 
observation of trends in various experimental results. The role of the 
quark-to-gluon ratio in the measured modification of fragmentation functions 
might be further explored by measurement of fragmentation at different 
pseudorapidities, measurement of fragmentation in $\gamma$-jet events as well 
as by a measurement of particle multiplicity inside jets. The multiplicity 
is clearly sensitive to the jet quenching~\cite{aadh:Salgado} but 
it is also powerful to distinguish between quark and gluon jets fragmenting in 
the vacuum~\cite{aaek:Gallicchio}.

One of striking observations we saw in the experimental results is the flatness 
of the jet nuclear modification factor at high-\pt\ reported by ATLAS and CMS. 
We have hinted at an importance of appropriate treatment of quark-to-gluon jet 
fractions to disentangle the features seen in the jet fragmentation. Different 
fractions of quark and gluon jets at different \pt\ and $\eta$ have to be 
clearly taken into account also when trying to understand the result of flat 
jet nuclear modification factor. The clear \pt\ and $\eta$ dependence of quark 
versus gluon fractions of partons initiating p+p jets might in fact help us to 
reveal some information about the difference in the quark and gluon jet 
quenching as we will illustrate. 
  The jet spectra can be generally described by the power-law distribution ($\mathrm{d}\sigma/\mathrm{d}\pt \sim \pt^k$). Let us imagine the
quenching as an effective shift of the spectra by $\Delta \pt$ at a given \pt\ value. Then, the \Raa\ can be
expressed as

\begin{equation}
\label{eqn:Raa}
\Raa = \frac{N_\mathrm{quark-jet}^\mathrm{quenched}(\pt) + N_\mathrm{gluon-jet}^\mathrm{quenched}(\pt)}
            {N_\mathrm{quark-jet}^\mathrm{vacuum}(\pt) + N_\mathrm{gluon-jet}^\mathrm{vacuum}(\pt)}
     = \frac{(\pt + \Delta\pt)^k f_q + (\pt + \kappa\Delta\pt)^k (1 - f_q) }{\pt^k}
\end{equation}
  where $\kappa$ is the effective color factor, $f_q=f_q(\pt,\eta)$ is the 
fraction of quark-to-inclusive jets, and exponent $k=-5$ for LHC 
energies~\cite{baah:ATLAS}. If we plug in standard QCD $\kappa = 9/4$, $f_q$ 
from Fig.\ref{fig:QtoG}, and \Raa\ of 0.5, we get for 100~GeV jet a reasonable 
$\Delta\pt$ of 11 GeV.
  It is clear that Eqn.\ref{eqn:Raa} represents
an over-simplification of the problem -- the whole dynamics of 
the quenching resides in the $\Delta\pt$. However, it allows us to see that the 
jet \Raa\ carries an important information which may be rigorously accessed if 
it is supplied by an information from another quark-to-gluon jet sensitive 
measurement such as appear to be the fragmentation\footnote{Especially the 
value in the minimum of $R_D(\xi)$ or $R_D(z)$ may be valuable for this 
purpose.}.

  This discussion illustrates that the kinematic coverage of the LHC data and 
the multitude of measurements summarized in this review give a very good 
discriminating power and ultimately should lead to a better understanding of the 
jet quenching.

%
%
%

\section{Summary and conclusions}

  In this short review we have summarized the up-to-date experimental results 
on jet suppression with high-\pt\ observables obtained from ALICE, ATLAS, and 
CMS.
 %
  The diversity of experimental measurements at high-\pt\ involving different 
objects (jets, charged particles, $\gamma$, $Z^0$), different observables (nuclear 
modification factors, fragmentation functions, high-\pt\ $v_2$, 
hadron-jet, $\gamma$-jet, or $Z^{0}$-jet correlations), and different kinematic 
selections provide clear possibilities to constrain the models of the jet 
quenching.
  The complexity of the physics problem (e.g. appropriate modeling of the space-time 
evolution of the medium along with the modeling of the jet quenching) and the 
complexity of the measurements (e.g. UE subtraction, fragmentation evaluated with respect to the 
quenched jet, possible influence of the UE by fragmentation) suggest a strong 
need of publicly available fully simulated models of the jet quenching. The 
full simulation allows for a validation of a given model by a precise 
comparison to variety of different experimental results in the same manner as 
different MC generators are confronted with the data in elementary collisions.

Experimental results on jets in heavy ion collisions at LHC clearly provide a 
rich set of data which has a great potential to shad light on the jet 
quenching, properties of the deconfined QCD medium, and fundamentals of the 
strong interaction. These results are expected to be followed by precision 
measurements after the upgrade of LHC.

~

{\it Disclosure statement:} The author is not aware of any affiliation, membership, 
funding, or financial holdings that might be perceived as affecting the 
objectivity of this review.

\section{Acknowledgment}

I'm grateful for stimulating discussions with Ji\v r\' i Dolej\v s\' i and 
Brain Cole. I'd like to thank to Aaron Angerami, Marco van Leeuwen, Alexander 
Milov, Mateusz Ploskon, and Christof Roland for a help with selecting 
appropriate data for the summary plots. This work was supported by grants 
MSM0021620859, UNCE 204020/2012, LG13009.


\begin{thebibliography}{0}

\bibitem{aaag:Karsch} F. Karsch, E. Laermann, Quark-Gluon Plasma III, ed. R. Hwa, \href{http://arxiv.org/abs/arXiv:hep-lat/0305025}{arXiv:hep-lat/0305025}.
\bibitem{aaad:Collins} J. C. Collins, M. J. Perry, \href{http://dx.doi.org/10.1103/PhysRevLett.34.1353}{Phys. Rev. Lett. {\bf 34} (1975) {\it 1353}}.
\bibitem{aaae:Cabbibo} N. Cabbibo, G. Parisi, \href{http://dx.doi.org/10.1016/0370-2693(75)90158-6}{Phys. Lett. {\bf B59} (1975) {\it 67}}.
\bibitem{aage:Shuryak} E. V. Shuryak, \href{http://dx.doi.org/10.1016/0370-2693(78)90370-2}{Phys. Lett. {\bf B78} (1978) {\it 150}}.
\bibitem{aaac:Bjorken} J. D. Bjorken, FERMILAB-PUB-82-059-THY (1982).
\bibitem{aaar:ALICE} K. Aamodt {\it et al.} [ALICE Collaboration], \href{http://dx.doi.org/10.1088/1748-0221/3/08/S08002}{JINST {\bf 3} (2008) {\it S08002}}.
\bibitem{baab:ATLAS} G. Aad {\it et al.} [ATLAS Collaboration], \href{http://dx.doi.org/10.1088/1748-0221/3/08/S08003}{JINST {\bf 3} (2008) {\it S08003}}.
\bibitem{aaaq:CMS} R. Adolphi {\it et al.} [CMS Collaboration], \href{http://dx.doi.org/10.1088/1748-0221/3/08/S08004}{JINST {\bf 3} (2008) {\it S08004}}.
\bibitem{aaap:BRAHMS} I. Arsene {\it et al.} [BRAHMS Collaboration], \href{http://dx.doi.org/10.1016/j.nuclphysa.2005.02.130}{Nucl. Phys. {\bf A757} (2005) {\it 1}}.
\bibitem{aaam:PHENIX} K. Adcox {\it et al.} [PHENIX Collaboration], \href{http://dx.doi.org/10.1016/j.nuclphysa.2005.03.086}{Nucl. Phys. {\bf A757} (2005) {\it 184-283}}.
\bibitem{aaao:PHOBOS} B. B. Back {\it et al.} [PHOBOS Collaboration], \href{http://dx.doi.org/10.1016/j.nuclphysa.2005.03.084}{Nucl. Phys. {\bf A757} (2005) {\it 28}}.
\bibitem{aaan:STAR} J. Adams {\it et al.} [STAR Collaboration], \href{http://dx.doi.org/10.1016/j.nuclphysa.2005.03.085}{Nucl. Phys. {\bf A757} (2005) {\it 102-183}}.
\bibitem{aagh:BraunMunzinger} P. Braun-Munzinger, J. Wambach, \href{http://dx.doi.org/10.1103/RevModPhys.81.1031}{Rev. Mod. Phys. {\bf 81} (2009) {\it 1031-1050}}.
\bibitem{aagc:Muller} B. Muller, J. L. Nagle, \href{http://dx.doi.org/10.1146/annurev.nucl.56.080805.140556}{Annu. Rev. Nucl. and Part. Phys. {\bf 56} (2006) {\it 93-135}}.
\bibitem{aagf:Jacobs} P. Jacobs, X.~N. Wang, \href{http://dx.doi.org/10.1016/j.ppnp.2004.09.001}{Prog. Part. Nucl. Phys. {\bf 54} (2005) {\it 443-534}}.
\bibitem{aagg:Huovinen} P. Huovinen, P. V. Ruuskanen, \href{http://dx.doi.org/10.1146/annurev.nucl.54.070103.181236}{Ann. Rev. Nucl. Part. Sci. {\bf 56} (2006) {\it 163-206}}.
\bibitem{aagd:Shuryak} E. V. Shuryak, \href{http://dx.doi.org/10.1016/j.nuclphysa.2004.10.022}{Nucl. Phys. {\bf A750} (2005) {\it 64-83}}.
\bibitem{aags:Accardi} A. Accardi, F. Arleo, W. K. Brooks, D. d'Enterria, V. Muccifora, \href{http://dx.doi.org/10.1393/ncr/i2009-10048-0}{Riv.Nuovo Cim. {\bf 32} (2010) {\it 439-553}}.
\bibitem{aafd:Muller} B. Muller, J. Schukraft, B. Wyslouch, \href{http://dx.doi.org/10.1146/annurev-nucl-102711-094910}{Annu. Rev. Nucl. Part. Sci. {\bf 62} (2012) {\it 361-386}}.
\bibitem{aafc:Singh} R. Singh, L. Kumar, P. Kumar Netrakanti, B. Mohanty, \href{http://arxiv.org/abs/arXiv:1304.2969}{arXiv:1304.2969} (unpublished).
\bibitem{aaca:Armesto} N. Armesto, B. Cole, C. Gale {\it et al.}, \href{http://dx.doi.org/10.1103/PhysRevC.86.064904}{Phys. Rev. {\bf C86} (2012) {\it 064904}}.
\bibitem{aafm:Bass} S. A. Bass, C. Gale, A. Majumder, C. Nonaka, G.-Y. Qin, T. Renk, J. Ruppert, \href{http://dx.doi.org/10.1103/PhysRevC.79.024901}{Phys. Rev. {\bf C79} (2009) {\it 024901}}.
\bibitem{aaai:Majumder} A. Majumder, M. Van Leeuwen, \href{http://dx.doi.org/10.1016/j.ppnp.2010.09.001}{Prog. Part. Nucl. Phys. {\bf 66} (2011) {\it 41-92}}.
\bibitem{aaah:Wiedemann} U. A. Wiedemann, Landolt-Boernstein Handbook of Physics, ed. R. Stock, \href{http://arxiv.org/abs/arXiv:0908.2306}{arXiv:0908.2306}.
\bibitem{aadj:MehtarTani} Y. Mehtar-Tani, J. G. Milhano, K. Tywoniuk, \href{http://dx.doi.org/10.1142/S0217751X13400137}{Int. J. of Mod. Phys. {\bf A28} (2013) {\it 1340013}}.
\bibitem{aahg:Baier} R. Baier, Yu-L. Dokshitzer, A. H. Mueller, S. Peigne, D. Schiff, \href{http://dx.doi.org/10.1016/S0550-3213(96)00553-6}{Nucl. Phys. {\bf B483} (1997) {\it 291-300}}.
\bibitem{aahb:Gyulassy} M. Gyulassy, P. Levai, I. Vitev, \href{http://dx.doi.org/10.1103/PhysRevLett.85.5535}{Phys. Rev. Lett. {\bf 85} (2000) {\it 5535}}.
\bibitem{aahc:Wang} X.-N. Wang, X. Guo, \href{http://dx.doi.org/10.1016/S0375-9474(01)01130-7}{Nucl. Phys. {\bf A696} (2001) {\it 788-832}}.
\bibitem{aahd:Zakharov} B.~G. Zakharov, \href{http://dx.doi.org/10.1134/1.1857266}{JETP Lett. {\bf 80} (2004) {\it 617-622}}.
\bibitem{aahe:Arnold} P.~B. Arnold, G.~D. Moore, L.~G. Yaffe, \href{http://dx.doi.org/10.1088/1126-6708/2001/12/009}{JHEP {\bf 0112} (2001) {\it 009}}.
\bibitem{aagj:Gyulassy} M. Gyulassy, Wang. X.-N, \href{http://dx.doi.org/10.1016/0550-3213(94)90079-5}{Nucl. Phys. {\bf B420} (1994) {\it 583}}.
\bibitem{aagk:Wang} X.-N. Wang, M. Gyulassy, M. Pl"umer, \href{http://dx.doi.org/10.1103/PhysRevD.51.3436}{Phys. Rev. {\bf D51} (1995) {\it 3436}}.
\bibitem{aagl:Mustafa} M. G. Mustafa, \href{http://dx.doi.org/10.1103/PhysRevC.72.014905}{Phys. Rev. {\bf C72} (2005) {\it 014905}}.
\bibitem{aaal:STAR} C. Adler {\it et al.} [STAR Collaboration], \href{http://dx.doi.org/10.1103/PhysRevLett.89.202301}{Phys. Rev. Lett. {\bf 89} (2002) {\it 202301}}.
\bibitem{aaak:PHENIX} K. Adcox {\it et al.} [PHENIX Collaboration], \href{http://dx.doi.org/10.1103/PhysRevLett.88.022301}{Phys. Rev. Lett. {\bf 88} (2002) {\it 022301}}.
\bibitem{aafb:STAR} C. Adler {\it et al.} [STAR Collaboration], \href{http://dx.doi.org/10.1103/PhysRevLett.90.082302}{Phys. Rev. Lett. {\bf 90} (2003) {\it 082302}}.
\bibitem{baaa:ATLAS} G. Aad {\it et al.} [ATLAS Collaboration], \href{http://dx.doi.org/10.1103/PhysRevLett.105.252303}{Phys. Rev. Lett. {\bf 105} (2010) {\it 252303}}.
\bibitem{aaab:CMS} S. Chatrchyan [CMS Collaboration], \href{http://dx.doi.org/10.1103/PhysRevC.84.024906}{Phys.~Rev. {\bf C84} (2011) {\it 024906}}.
\bibitem{baam:ATLAS} ATLAS Collaboration, ATLAS-CONF-2011-075 (2011).
\bibitem{aabk:CMS} S. Chatrchyan {\it et al.} [CMS Collaboration], \href{http://dx.doi.org/10.1016/j.physletb.2012.04.058}{Phys. Lett. B {\bf B712} (2012) {\it 176-197}}.
\bibitem{aacb:Qin} G. Y. Qin, B. Muller, \href{http://dx.doi.org/10.1103/PhysRevLett.106.162302}{Phys.Rev.Lett. {\bf 106} (162302) {\it 2011}}.
\bibitem{aace:Young} C. Young, B. Schenke, S. Jeon, C. Gale, \href{http://dx.doi.org/10.1103/PhysRevC.84.024907}{Phys. Rev. {\bf C84} (2011) {\it 024907}}.
\bibitem{aacf:Renk} T. Renk, \href{http://dx.doi.org/10.1103/PhysRevC.86.061901}{Phys. Rev. {\bf C86} (2012) {\it 061901}}.
\bibitem{aade:He} Y. He, I. Vitev, B.-W. Zhang, \href{http://dx.doi.org/10.1016/j.physletb.2012.05.054}{Phys. Lett. {\bf B713} (2012) {\it 224-232}}.
\bibitem{aadg:CasalderreySolana} J. Casalderrey-Solana, J. G. Milhano, U. A. Wiedemann, \href{http://dx.doi.org/10.1088/0954-3899/38/3/035006}{J. Phys. {\bf G38} (2011) {\it 035006}}.
\bibitem{aahk:Lokhtin} I. P. Lokhtin, A. V. Belyaev, A. M. Snigirev, \href{http://dx.doi.org/10.1140/epjc/s10052-011-1650-1}{Eur. Phys. J. {\bf C71} (2011) {\it 1650}}.
\bibitem{aaba:Cacciari} M. Cacciari, G. P. Salam, G. Soyez, \href{http://dx.doi.org/10.1088/1126-6708/2008/04/063}{JHEP {\bf 0804} (063) {\it 2008}}.
\bibitem{aabf:CMS} CMS Collaboration, CMS-PAS-PFT-09-001 (2009).
\bibitem{aahn:ALICE} R. Reed {\it et al.} [ALICE Collaboration], \href{http://arxiv.org/abs/arXiv:1304.5945}{arXiv:1304.5945} (unpublished).
\bibitem{aabb:Cacciari} M. Cacciari, G. P. Salam, G. Soyez, \href{http://dx.doi.org/10.1088/1126-6708/2008/04/005}{JHEP {\bf 0804} (2008) {\it 005}}.
\bibitem{aabc:Cacciari} M. Cacciari, G. P. Salam, \href{http://dx.doi.org/10.1016/j.physletb.2007.09.077}{Phys. Lett. {\bf B659} (2008) {\it 119}}.
\bibitem{aabd:Cacciari} M. Cacciari, J. Rojo, G. P. Salam, G. Soyez, \href{http://dx.doi.org/10.1140/epjc/s10052-011-1539-z}{Eur. Phys. J. {\bf C71} (2011) {\it 1539}}.
\bibitem{baah:ATLAS} G. Aad {\it et al.} [ATLAS Collaboration], \href{http://dx.doi.org/10.1016/j.physletb.2013.01.024}{Phys. Lett. {\bf B719} (2013) {\it 220-241}}.
\bibitem{aada:ALICE} B. Abelev {\it et al.} [ALICE Collaboration], \href{http://dx.doi.org/10.1007/JHEP03(2012)053}{JHEP {\bf 03} (2012) {\it 053}}.
\bibitem{aabe:Cacciari} M. Cacciari, G. P. Salam, G. Soyez, \href{http://dx.doi.org/10.1140/epjc/s10052-011-1692-4}{Eur. Phys. J. {\bf C71} (2011) {\it 1692}}.
\bibitem{aabq:CMS} CMS Collaboration, PAS-HIN-12-004 (2012).
\bibitem{aads:CMS} M. Nguyen {\it et al.} [CMS Collaboration], \href{http://dx.doi.org/10.1088/0954-3899/38/12/124151 }{J. Phys. {\bf G38} (2011) {\it 124151}}.
\bibitem{aahm:ALICE} B. Abelev {\it et al.} [ALICE Collaboration], \href{http://dx.doi.org/10.1016/j.physletb.2013.04.026}{Phys. Lett. {\bf B722} (2013) {\it 262}}.
\bibitem{aaat:Sjostrand} T. Sjostrand, S. Mrenna, P. Skands, \href{http://dx.doi.org/10.1088/1126-6708/2006/05/026}{JHEP {\bf 05} (2006) {\it 026}}.
\bibitem{aaav:Wang} X. N. Wang, M. Gyulassy, \href{http://dx.doi.org/10.1016/0010-4655(94)90057-4}{Comput. Phys. Commun. {\bf 83} (1994) {\it 307}}.
\bibitem{aaaw:Lokhtin} I. P. Lokhtin, A. M. Snigirev, \href{http://dx.doi.org/10.1140/epjc/s2005-02426-3}{Eur. Phys. J. {\bf C45} (2006) {\it 211}}.
\bibitem{aagv:STAR} P.~M. Jacobs {\it et al.} [STAR Collaboration], \href{http://arxiv.org/abs/arXiv:1012.2406}{arXiv:1012.2406} (unpublished).
\bibitem{aadu:Apolinario} L. Apolinario, N. Armesto, L. Cunqueiro, \href{http://dx.doi.org/10.1007/JHEP02(2013)022}{JHEP {\bf 1302} (2013) {\it 022}}.
\bibitem{aabg:CMS} CMS Collaboration, CMS-PAS-PFT-10-002 (2010).
\bibitem{baan:ATLAS} ATLAS Collaboration, ATLAS-CONF-2012-045 (2012).
\bibitem{aabu:Fabjan} W. C. Fabjan, F. Gianotti, \href{http://dx.doi.org/10.1103/RevModPhys.75.1243}{Rev. Mod. Phys. {\bf 75} (2003) {\it 1243-1286}}.
\bibitem{babf:ATLAS} M. Spousta {\it et al.} [ATLAS Collaboration], \href{http://dx.doi.org/10.1088/1742-6596/270/1/012013}{J. Phys. Conf. Ser. 270 {\bf 270} (2011) {\it 012013}}.
\bibitem{aaho:Miller} M. L. Miller, K. Reygers, S. J. Sanders, P. Steinberg, \href{http://dx.doi.org/10.1146/annurev.nucl.57.090506.123020}{Ann. Rev. Nucl. Part. Sci. {\bf 57} (2007) {\it 205}}.
\bibitem{aadv:ALICE} B. Abelev {\it et al.} [ALICE Collaboration], \href{http://dx.doi.org/10.1016/j.physletb.2013.01.051}{Phys. Lett. {\bf B720} (2013) {\it 52-62}}.
\bibitem{baal:ATLAS} ATLAS Collaboration, ATLAS-CONF-2011-079 (2011).
\bibitem{aabj:CMS} S. Chatrchyan {\it et al.} [CMS Collaboration], \href{http://dx.doi.org/10.1140/epjc/s10052-012-1945-x}{Eur. Phys. J. {\bf C72} (2012) {\it 1945}}.
\bibitem{aabs:CasalderreySolana} J. Casalderrey-Solana, A. Milov, \href{http://arxiv.org/abs/arXiv:1210.8271}{arXiv:1210.8271} (unpublished).
\bibitem{babd:ATLAS} ATLAS Collaboration, \href{http://dx.doi.org/10.1103/PhysRevD.83.052003}{Phys. Rev. {\bf D83} (2011) {\it 052003}}.
\bibitem{aafl:Cole} B. A. Cole, \href{http://dx.doi.org/10.1016/j.nuclphysa.2006.06.043}{Nucl. Phys. {\bf A774} (2006) {\it 225-236}}.
\bibitem{aahj:Renk} T. Renk, \href{http://arxiv.org/abs/arXiv:1302.3710 }{arXiv:1302.3710 } (unpublished).
\bibitem{aafk:Abreu} S. Abreu, S. V. Akkelin, J. Alam {\it et al.}, \href{http://dx.doi.org/10.1088/0954-3899/35/5/054001}{J. Phys. {\bf G35} (2008) {\it 054001}}.
\bibitem{aahl:Renk} T. Renk, \href{http://arxiv.org/abs/arXiv:1207.4885}{arXiv:1207.4885} (unpublished).
\bibitem{aago:ALICE} K. Aamond {\it et al.} [ALICE Collaboration], \href{http://dx.doi.org/10.1103/PhysRevLett.108.092301}{Phys. Rev. Lett. {\bf 108} (2012) {\it 092301}}.
\bibitem{aabr:CMS} CMS Collaboration, PAS-HIN-12-010 (2012).
\bibitem{aagr:Pantuev} V. S. Pantuev, \href{http://dx.doi.org/10.1134/S0021364007020026}{JETP Lett. {\bf 85} (2007) {\it 104-108}}.
\bibitem{aahf:Renk} T. Renk, K. Eskola, \href{http://dx.doi.org/10.1103/PhysRevC.75.054910}{Phys.Rev. {\bf C75} (2007) {\it 054910}}.
\bibitem{aabl:CMS} S. Chatrchyan {\it et al.} [CMS Collaboration], \href{http://dx.doi.org/10.1103/PhysRevLett.109.022301}{Phys. Rev. Lett. {\bf 109} (2012) {\it 022301}}.
\bibitem{baau:ATLAS} ATLAS Collaboration, ATLAS-CONF-2012-116 (2012).
\bibitem{aafu:Baier} R. Baier, Y. L. Dokshitzer, A. H. Mueller, D. Schiff, \href{http://dx.doi.org/10.1103/PhysRevC.58.1706}{Phys. Rev. {\bf C58} (1998) {\it 1706-1713}}.
\bibitem{aafs:Peigne} S. Peigne, A. V. Smilga, \href{http://dx.doi.org/10.3367/UFNe.0179.200907a.0697}{Phys. Usp. {\bf 52} (2009) {\it 659}}.
\bibitem{aafo:Gubser} S. S. Gubser, D. R. Gulotta, S. P. Silviu, F. D. Rocha, \href{http://dx.doi.org/10.1088/1126-6708/2008/10/052}{JHEP {\bf 0810} (2008) {\it 052}}.
\bibitem{aacr:Wicks} S. Wicks, W. Horowitz, M. Djordjevic, M. Gyulassy, \href{http://dx.doi.org/10.1016/j.nuclphysa.2006.12.048}{Nucl. Phys. {\bf A784} (2007) {\it 426-442}}.
\bibitem{aaga:Marquet} C. Marquet, T. Renk, \href{http://dx.doi.org/10.1016/j.physletb.2010.01.076}{Phys. Lett. {\bf B685} (2010) {\it 270-276}}.
\bibitem{aafv:Renk} T. Renk, \href{http://dx.doi.org/10.1103/PhysRevC.83.024908}{Phys. Rev. {\bf C83} (2011) {\it 024908}}.
\bibitem{aafy:Betz} B. Betz, M. Gyulassy, G. Torrieri, \href{http://dx.doi.org/10.1103/PhysRevC.84.024913}{Phys. Rev. {\bf C84} (2011) {\it 024913}}.
\bibitem{aafw:Jia} J. Jia, W. A. Horowitz, J. Liao, \href{http://dx.doi.org/10.1103/PhysRevC.84.034904}{Phys. Rev. {\bf C84} (2011) {\it 034904}}.
\bibitem{aafp:Liu} H. Liu, K. Rajagopal, U. A. Wiedemann, \href{http://dx.doi.org/10.1103/PhysRevLett.97.182301}{Phys. Rev. Lett. {\bf 97} (2006) {\it 182301}}.
\bibitem{aagb:Gubser} S. S. Gubser, A. Karch, \href{http://dx.doi.org/10.1146/annurev.nucl.010909.083602}{Ann. Rev. Nucl. Part. Sci. {\bf 59} (2009) {\it 145-168}}.
\bibitem{aacg:PHENIX} A. Adare {\it et al.} [PHENIX Collaboration], \href{http://dx.doi.org/10.1103/PhysRevLett.105.142301}{Phys. Rev. Lett. {\bf 105} (2010) {\it 142301}}.
\bibitem{aacd:Betz} B. Betz, \href{http://dx.doi.org/10.1140/epja/i2012-12164-8}{Eur. Phys. J. {\bf A48} (2012) {\it 164}}.
\bibitem{aagm:ALICE} B. Abelev {\it et al.} [ALICE Collaboration], \href{http://dx.doi.org/10.1016/j.physletb.2012.12.066}{Phys. Lett. {\bf B719} (2013) {\it 18-28}}.
\bibitem{baae:ATLAS} G. Aad {\it et al.} [ATLAS Collaboration], \href{http://dx.doi.org/10.1016/j.physletb.2011.12.056}{Phys. Lett. {\bf B707} (2012) {\it 330-348}}.
\bibitem{aagn:ALICE} K. Aamodt {\it et al.} [ALICE Collaboration], \href{http://dx.doi.org/10.1103/PhysRevLett.105.252302}{Phys. Rev. Lett. {\bf 105} (2010) {\it 252302}}.
\bibitem{aaft:CMS} S. Chatrchyan {\it et al.} [CMS Collaboration], \href{http://dx.doi.org/10.1103/PhysRevC.87.014902}{Phys. Rev. C 87 (2013) 014902 {\bf C87} (2013) {\it 014902}}.
\bibitem{aaaj:Wang} X. N. Wang, \href{http://dx.doi.org/10.1016/j.physletb.2003.11.011}{Phys. Lett. {\bf B579} (2004) {\it 299-308}}.
\bibitem{baar:ATLAS} ATLAS Collaboration, ATLAS-CONF-2012-052 (2012).
\bibitem{aabi:CMS} S. Chatrchyan {\it et al.} [CMS Collaboration], \href{http://dx.doi.org/10.1016/j.physletb.2012.02.077}{Phys. Lett. {\bf B710} (2012) {\it 256-277}}.
\bibitem{baai:ATLAS} G. Aad {\it et al.} [ATLAS Collaboration], \href{http://dx.doi.org/10.1103/PhysRevLett.110.022301}{Phys. Rev. Lett. {\bf 110} (2013) {\it 022301}}.
\bibitem{baak:ATLAS} ATLAS Collaboration, ATLAS-CONF-2011-078 (2011).
\bibitem{aabh:CMS} S. Chatrchyan {\it et al.} [CMS Collaboration], \href{http://dx.doi.org/10.1103/PhysRevLett.106.212301}{Phys. Rev. Lett. {\bf 106} (2011) {\it 212301}}.
\bibitem{aafn:Paukkunen} H. Paukkunen, C. A. Salgado, \href{http://dx.doi.org/10.1007/JHEP03(2011)071}{JHEP {\bf 1103} (2011) {\it 071}}.
\bibitem{aadp:Wang} X.-N. Wang, Z. Huang, I. Sarcevic, \href{http://dx.doi.org/10.1103/PhysRevLett.77.231}{Phys. Rev. Lett. {\bf 77} (1996) {\it 231-234}}.
\bibitem{babb:ATLAS} ATLAS Collaboration, ATLAS-CONF-2012-121 (2012).
\bibitem{baay:ATLAS} ATLAS Collaboration, ATLAS-CONF-2012-119 (2012).
\bibitem{aabo:CMS} S. Chatrchyan {\it et al.} [CMS Collaboration], \href{http://dx.doi.org/10.1016/j.physletb.2012.11.003}{Phys. Lett. {\bf B718} (2013) {\it 773-794}}.
\bibitem{aado:Dai} W. Dai, I. Vitev, B.-W. Zhang, \href{http://dx.doi.org/10.1103/PhysRevLett.110.142001}{Phys. Rev. Lett. {\bf 110} (2013) {\it 142001}}.
\bibitem{babe:ATLAS} ATLAS Collaboration, ATL-PHYS-PUB-2012-002 (2012).
\bibitem{aafa:CMS} CMS Collaboration, \href{https://indico.cern.ch/abstractDisplay.py/getAttachedFile?abstractId=144&resId=3&confId=175067}{Future Heavy-Ion Programme at CMS (Contribution to the Update of the European Strategy for Particle Physics)}.
\bibitem{aacj:Dokshitzer} Yu. L. Dokshitzer, D. E. Kharzeev, \href{http://dx.doi.org/10.1016/S0370-2693(01)01130-3}{Phys. Lett. {\bf B519} (2001) {\it 199-206}}.
\bibitem{aack:PHENIX} A. Adare {\it et al.} [PHENIX Collaboration], \href{http://dx.doi.org/10.1103/PhysRevLett.98.172301}{Phys. Rev. Lett. {\bf 98} (2007) {\it 172301}}.
\bibitem{aacl:STAR} B. I. Abelev {\it et al.} [STAR Collaboration], \href{http://dx.doi.org/10.1103/PhysRevLett.98.192301}{Phys. Rev. Lett. {\bf 98} (2007) {\it 192301}}.
\bibitem{aacm:PHENIX} A. Adare {\it et al.} [PHENIX Collaboration], \href{http://dx.doi.org/10.1103/PhysRevC.84.044905}{Phys. Rev. {\bf C84} (2011) {\it 044905}}.
\bibitem{aacn:Djordjevic} M. Djordjevic, \href{http://dx.doi.org/10.1103/PhysRevC.85.034904}{Phys. Rev. {\bf C85} (2012) {\it 034904}}.
\bibitem{aaco:Gossiaux} P. B. Gossiaux, J. Aicheli, \href{http://dx.doi.org/10.1103/PhysRevC.78.014904}{Phys. Rev. {\bf C78} (2008) {\it 014904}}.
\bibitem{aacp:Uphoff} J. Uphoff, O. Fochler, Z. Xu, C. Greiner, \href{http://dx.doi.org/10.1103/PhysRevC.84.024908}{Phys. Rev. {\bf C84} (2011) {\it 024908}}.
\bibitem{aadm:Horowitz} W. A. Horowitz, M. Gyulassy, \href{http://dx.doi.org/10.1088/0954-3899/35/10/104152}{J. Phys. {\bf G35} (2008) {\it 104152}}.
\bibitem{aadl:VanHees} H. Van Hees, M. Mannarelli, V. Greco, R. Rapp, \href{http://dx.doi.org/10.1103/PhysRevLett.100.192301}{Phys. Rev. Lett. {\bf 100} (2008) {\it 192301}}.
\bibitem{aadk:Adil} A. Adil, I. Vitev, \href{http://dx.doi.org/10.1016/j.physletb.2007.03.050}{Phys. Lett. {\bf B649} (2007) {\it 139-146}}.
\bibitem{aabt:CMS} CMS Collaboration, PAS-HIN-12-003 (2012).
\bibitem{baap:ATLAS} ATLAS Collaboration, ATLAS-CONF-2012-050 (2012).
\bibitem{aagp:ALICE} B. Abelev {\it et al.} [ALICE Collaboration], \href{http://dx.doi.org/10.1103/PhysRevLett.109.112301}{Phys. Rev. Lett. {\bf 109} (2012) {\it 112301}}.
\bibitem{aagt:CMS} S. Chatrchyan {\it et al.} [CMS Collaboration], \href{http://dx.doi.org/10.1007/JHEP05(2012)063}{JHEP {\bf 1205} (2012) {\it 063}}.
\bibitem{aadh:Salgado} C. A. Salgado, U. A. Wiedemann, \href{http://dx.doi.org/10.1103/PhysRevLett.93.042301}{Phys. Rev. Lett. {\bf 93} (2004) {\it 042301}}.
\bibitem{aadc:Lokhtin} I. P. Lokhtin, A. M. Snigirev, \href{http://dx.doi.org/10.1016/j.physletb.2003.06.013}{Phys.Lett. B567 (2003) 39-45 {\bf B567} (2003) {\it 39-45}}.
\bibitem{aadw:Borghini} N. Borghini, U. A. Wiedemann, \href{http://arxiv.org/abs/arXiv:hep-ph/0506218}{arXiv:hep-ph/0506218} (unpublished).
\bibitem{aacu:Armesto} N. Armesto, L. Cunqueiro, C. A. Salgado, W.-C. Xiang, \href{http://dx.doi.org/10.1088/1126-6708/2008/02/048}{JHEP {\bf 0802} (2008) {\it 048}}.
\bibitem{aadd:Vitev} I. Vitev, S. Wicks, B.-W. Zhang, \href{http://dx.doi.org/10.1088/1126-6708/2008/11/093}{JHEP {\bf 11} (2008) {\it 093}}.
\bibitem{aady:Armesto} N. Armesto, L. Cunqueiro, C. A. Salgado, \href{http://dx.doi.org/10.1140/epjc/s10052-009-1133-9}{Eur. Phys. J. {\bf C63} (2009) {\it 679-690}}.
\bibitem{aadz:Renk} T. Renk, \href{http://dx.doi.org/10.1103/PhysRevC.79.054906}{Phys. Rev. {\bf C79} (2009) {\it 054906}}.
\bibitem{aabn:CMS} S. Chatrchyan {\it et al.} [CMS Collaboration], \href{http://dx.doi.org/10.1007/JHEP10(2012)087}{JHEP {\bf 10} (2012) {\it 087}}.
\bibitem{aabp:CMS} CMS Collaboration, PAS-HIN-12-013 (2012).
\bibitem{baat:ATLAS} ATLAS Collaboration, ATLAS-CONF-2012-115 (2012).
\bibitem{aadq:Zapp} K. C. Zapp, F. Krauss, U. A. Wiedemann, \href{http://dx.doi.org/10.1007/JHEP03(2013)080}{JHEP {\bf 1303} (2013) {\it 080}}.
\bibitem{aadr:Kharzeev} D. E. Kharzeev, F. Loshaj, \href{http://arxiv.org/abs/arXiv:1212.5857}{arXiv:1212.5857} (unpublished).
\bibitem{aaei:ALEPH} D. Buskulic {\it et al.} [ALEPH Collaboration], \href{http://dx.doi.org/10.1016/0370-2693(96)00849-0}{Phys. Lett. {\bf B384} (1996) {\it 353-364}}.
\bibitem{aaej:OPAL} G. Alexander {\it et al.} [OPAL Collaboration], \href{http://dx.doi.org/10.1007/s002880050059}{Z. Phys. C69 {\bf C69} (1996) {\it 543-560}}.
\bibitem{aaef:DeFlorian} D. De Florian, R. Sassot, M. Stratmann, \href{http://dx.doi.org/10.1103/PhysRevD.76.074033}{Phys. Rev. {\bf D76} (2007) {\it 074033}}.
\bibitem{aaeg:CDF} D. Acosta {\it et al.} [CDF Collaboration], \href{http://dx.doi.org/10.1103/PhysRevD.71.112002}{Phys. Rev. {\bf D71} (2005) {\it 112002}}.
\bibitem{aaeh:ALEPH} R. Barate {\it et al.} [ALEPH Collaboration], \href{http://dx.doi.org/10.1007/s100520000474}{Eur. Phys. J. {\bf C17} (2008) {\it 1-18}}.
\bibitem{aaek:Gallicchio} J. Gallicchio, M. D. Schwartz, \href{http://dx.doi.org/10.1007/JHEP04(2013)090}{JHEP {\bf 1304} (2013) {\it 090}}.

\end{thebibliography}
\end{document}